\documentclass[
12pt,
superscriptaddress,
showpacs, 
showkeys,
amsmath,amssymb,
aip,
jcp,
floatfix
]{revtex4-1}

\usepackage{hyperref}
\usepackage{graphicx}
\usepackage{amsmath}
\usepackage{amsfonts}
\usepackage{amssymb}
\usepackage{yfonts}
\usepackage{bm}
\usepackage{siunitx}
\usepackage{enumitem}
\usepackage{color}
\usepackage{ulem}

\makeatletter
\newcommand*{\centerfloat}{%
  \parindent \z@
  \leftskip \z@ \@plus 1fil \@minus \textwidth
  \rightskip\leftskip
  \parfillskip \z@skip}
\makeatother

\setlist[itemize]{noitemsep, topsep=0pt}

\global\long\def\V#1{\boldsymbol{#1}}   
\global\long\def\T#1{\boldsymbol{#1}}   
\global\long\def\D#1{\Delta#1}          
\newcommand{\paren}[1]{{(#1)}}          
\newcommand{\grad}{\V{\nabla}}          
\newcommand{\divg}{\V{\nabla}\cdot}     
\newcommand{\curl}{\V{\nabla}\times}    
\newcommand{\lapl}{\nabla^2}            

\begin{document}

\title{Molecular Hydrodynamics: Vortex Formation and Sound Wave Propagation}

\author{Kyeong Hwan Han}
\affiliation{Department of Chemistry, Korea Advanced Institute of Science and Technology (KAIST), Daejeon 34141, Republic of Korea}

\author{Changho Kim}
\email{changhokim@lbl.gov}
\affiliation{Computational Research Division, Lawrence Berkeley National Laboratory, Berkeley, California 94720, USA}

\author{Peter Talkner}
\affiliation{Institut f\"ur Physik, Universit\"at Augsburg, 86159 Augsburg, Germany}

\author{George Em Karniadakis}
\affiliation{Division of Applied Mathematics, Brown University, Providence, Rhode Island 02912, USA}

\author{Eok Kyun Lee}
\email{eklee@kaist.ac.kr}
\affiliation{Department of Chemistry, Korea Advanced Institute of Science and Technology (KAIST), Daejeon 34141, Republic of Korea}

\date{\today}

\begin{abstract}
In the present study, quantitative feasibility tests of the hydrodynamic description of a two-dimensional fluid at the molecular level are performed, both with respect to length and time scales.
Using high-resolution fluid velocity data obtained from extensive molecular dynamics simulations, we computed the transverse and longitudinal components of the velocity field by the Helmholtz decomposition and compared them with those obtained from the linearized Navier--Stokes (LNS) equations with time-dependent transport coefficients.
By investigating the vortex dynamics and the sound wave propagation in terms of these field components, we confirm the validity of the LNS description for times comparable to or larger than several mean collision times.
The LNS description still reproduces the transverse velocity field accurately at smaller times, but it fails to predict characteristic patterns of molecular origin visible in the longitudinal velocity field.
Based on these observations, we validate the main assumptions of the mode-coupling approach.
The assumption that the velocity autocorrelation function can be expressed in terms of the fluid velocity field and the tagged particle distribution is found to be remarkably accurate even for times comparable to or smaller than the mean collision time.
This suggests that the hydrodynamic-mode description remains valid down to the molecular scale.
\end{abstract}

\maketitle

\section{\label{sec_intro}Introduction}

In principle, the hydrodynamics behavior of a fluid is determined by the dynamics of the molecules constituting the considered fluid.
In practice, however, direct molecular dynamics (MD) simulations of hydrodynamic phenomena are severely limited in time and become prohibitive as the system size goes beyond the micrometer scale.
The difficulty originates from the large scale difference between MD and hydrodynamics.
To resolve the collisions between fluid molecules, the time step and the shortest length scales of MD simulations must resolve the mean collision time and the size of a fluid molecule, respectively.
On the other hand, hydrodynamics describes the collective motion of fluid molecules, which corresponds to the zero-wavenumber limit.
In this limit, density fields of the globally conserved quantities such as mass, energy, and momentum are only considered.~\cite{HansenMcDonald2013, BoonYip1980, BalucaniZoppi1995, Forster1975}
Nevertheless, the MD simulation technique~\cite{AllenTildesley1989, FrenkelSmit2001, Rapaport2004} represents an indispensable tool to investigate the molecular origins of the hydrodynamics theory from first principles.
Powerful computing capacities available now are decisive in bridging the gap between the continuum-based analysis of transport phenomena and their modeling on the microscopic level.~\cite{Rapaport2014, KarniadakisBeskokAluru2005, MareschalHolian1992}

There have been various attempts to understand the hydrodynamic description from the molecular viewpoint by means of the MD simulation technique.
The first approach was to compute transport coefficients appearing in the phenomenological hydrodynamic description.
It is based on the Green--Kubo relations, where each transport coefficient is expressed as the time integral of the autocorrelation function of a  corresponding dynamical variable.
While this approach provides a practical methodology for estimating transport coefficients,~\cite{AllenTildesley1989, FrenkelSmit2001, Rapaport2004} it assumes the validity of the phenomenological hydrodynamic description and the complete scale separation between MD and hydrodynamics.

One question raised in various MD studies is whether the hydrodynamic description still holds down to the molecular scale, where the validity of the continuum description becomes questionable.
To this end, some well-known results from hydrodynamics have been tested in a molecular setting.
For example, the applicability of the Stokes--Einstein relation for a molecular-sized tracer particle was extensively investigated.
While overall good agreement of MD simulation results with the Stokes--Einstein relation was reported,~\cite{Ould-KaddourLevesque2000, CappelezzoCapellariPezzinCoelho2007, OhtoriIshii2015} this does not give a definite answer to the original question due to the subtlety existing in determining the particle-solvent boundary condition and the hydrodynamic radius. 

Moreover, specifically nonlinear aspects of fluid dynamics have been studied by means of MD simulations such as the formation of eddies behind an obstacle,~\cite{RapaportClementi1986, CuiEvans1992, SunHeLiTao2010} structure formation and flow instabilities,~\cite{MareschalMansourPuhlKestemont1988, Rapaport1988,HirshfeldRapport1998, Rapaport2006, TrevelyanZaki2016} flow of immiscible fluids,~\cite{KoplikBaravarWillemsen1988,KadauGermannHadjiconstantinouEtAl2004} and turbulent mixing,~\cite{DzwinelAldaPogodaYuen2000} to name but a few.
The emerging flow patterns suggest that even at the considered molecular length scales the respective systems can be treated as continuum fluids.
Points of instability, such as bifurcation points, however, are prone to large fluctuations seen in the MD results rendering a quantitative comparison with the hydrodynamic description difficult.~\cite{Rapaport2014, Rapaport1992}    

Another direction of investigation is to understand the time correlation functions of a molecular fluid at equilibrium by using the hydrodynamic theory.
Since these quantities are directly related to the transport properties, one can investigate the continuous transition from the molecular time scale to the hydrodynamic time scale.
The best-known and most extensively studied quantity is the velocity autocorrelation function (VACF) of a tagged particle in a $d$-dimensional simple fluid ($d=2$, 3).
Alder and Wainwright observed an  algebraic decay proportional to $t^{-d/2}$ and showed that this slow decay is caused by a vortex flow forming around the particle.~\cite{AlderWainwright1970}
Several types of theoretical approaches, having a variety of theoretical perspectives and mathematical sophistications, have been applied and refined to account for this unexpected behavior.
For example, the mode-coupling theory~\cite{PomeauResibois1975} relates the algebraic decay of the VACF to the vorticity diffusion based on the hydrodynamic description, whereas the kinetic theory~\cite{Cohen1993} sees its origin as correlated binary collisions (i.e., ring collisions) in the microscopic dynamics.
Nevertheless, all approaches give the identical expression for the long-time decay.
Recent extensive MD studies have confirmed that the emergence of the long-time tail is universal for fluid systems.~\cite{McDonoughRussoSnook2001, DibOuld-KaddorLevesque2006, Isobe2008, LesnickiVuilleumierCarofRotenberg2016}

The validity of the assumptions inherent in the derivation of governing equations of hydrodynamics  and their limitations have not been systematically scrutinized yet by MD simulations.
In particular, no general rules are known indicating the actual ranges of validity of the common restriction to large temporal and spatial scales for specific characteristics, say, of a flow pattern.
Also, in general, one does not know precisely when deviations from the hydrodynamic picture must be expected and of which nature and magnitude they would be.  
A further fundamental postulate of hydrodynamics is that of local equilibrium. It presupposes the separation of temporal and spatial scales. 
  
The flux correlation functions determining the transport coefficients via the Green--Kubo relations, such as viscosity, heat conduction, and diffusion coefficients, typically display a rapid decay at short time scales characteristic of the molecular motion and a slowly decaying long-time tail resulting from relatively large scale spatial structures of hydrodynamic nature.
While in three dimensions the total decay rate, including the hydrodynamically caused slow contribution, is sufficiently fast to yield finite, time-independent transport coefficients, the relevant hydrodynamic patterns in a two-dimensional fluid are more persistent.
Their decay is so slow that, for example, the diffusion coefficient characterizing the spreading of a tagged particle diverges.

In this paper, we study the hydrodynamic description of collective modes in a molecular fluid system by examining the mode-coupling approach.~\cite{ErnstHaugeLeeuwen1971b}
To this end, we performed an MD simulation study of a two-dimensional simple fluid by computing relevant hydrodynamic modes directly from MD simulations.
In order to determine the flow pattern that is generated by the motion of a tagged particle, we select those initial configurations from the equilibrium distribution for which the tagged particle sits in the origin and moves in the positive $x$-direction.
Letting all particles interact via short-range pairwise potentials, a flow pattern emerges that is constructed from the positions and velocities of all particles resulting from the MD simulation.
The resulting flow pattern can be decomposed into a potential and a solenoidal flow by means of the Helmholtz decomposition.
The latter flow corresponds to a vortex pattern as it was predicted by Alder and Wainwright~\cite{AlderWainwright1970} and confirmed in \cite{Isobe2009, HoefFrenkelLadd1991}.
The two flow fields are compared to the solutions of the linearized Navier--Stokes (LNS) equations containing time-dependent transport coefficients resulting from Green--Kubo-like formulas with the actual time as upper integration limit.  
The spatial resolution of the MD-based flow patterns is at 10\% of the fluid particle diameter, considerably finer than the 70\% resolution used in the recent work.~\cite{Isobe2009}
Such high resolution is needed in order to obtain a reliable separation into the potential and the solenoidal flow patterns. 
This though requires a large ensemble of MD data in order to keep the statistical errors sufficiently small. 
For efficient computation, we employed not only the ensemble average (over independent samples) but also an average over all particles and a large number of instants of time.

The rest of the paper is organized as follows.
In Section~\ref{sec_background}, the necessary theory is reviewed.
In Section~\ref{sec_num_proc}, our MD model and the numerical procedures are introduced.
In Section~\ref{sec_res}, the MD simulation results are presented and analyzed.  
Section~\ref{sec_summary} concludes the paper with a summary and an outlook.

\section{\label{sec_background}Background}

From the viewpoint of the mode-coupling approach, to obtain the VACF of a tagged particle, two field variables, the fluid velocity and the distribution of the tagged particle, are essential.
While a similar argument can be found somewhere else (e.g., \cite{DibOuld-KaddorLevesque2006, Beijeren1982}), we derive a \textit{refined} expression~\eqref{Ctlongtime} for the long-time VACF by assuming time-dependent transport coefficients.
This leads to a substantially improved agreement with MD simulation results for two-dimensional fluids.
On the other hand, with this refined approach, one recovers the well-known expression~\cite{ErnstHaugeLeeuwen1971b} for the case of constant coefficients.
In Section~\ref{subsec_condvel1}, we derive relations between the VACF and the average velocity of the tagged particle conditioned on its initial velocity.
In Section~\ref{subsec_condvel2}, we obtain an approximated form of the latter in terms of a particular fluid velocity field and the tagged particle distribution. 
In Section~\ref{subsec_hydro}, we obtain the VACF with the help of the hydrodynamic forms of these fields.

\subsection{\label{subsec_condvel1}VACF and Conditional Average Velocity}

We consider a tagged particle suspended in an isotropic fluid in $d$ dimensions; for detailed description on the molecular setting, see Section~\ref{subsec_sys_MD}.
By denoting the velocity of the tagged particle at the time $t$ by $\V{v}(t)$, we can express the VACF, $C(t)=\langle\V{v}(0)\cdot\V{v}(t)\rangle$, in terms of the average velocity at time $t$, $\langle\V{v}(t)|\V{v}_0\rangle$, conditioned on the initial velocity $\V{v}(0) \equiv \V{v}_0$, as
\begin{equation}
\label{VACFrel1}
C(t)=\Big< \V{v}_0\cdot\langle\V{v}(t)|\V{v}_0\rangle\Big>,
\end{equation}
whereby the outer average is determined by the Maxwell--Boltzmann velocity distribution $\Phi(\V{v}_0)$.

Because of the isotropy of the velocity in equilibrium, the conditional average points into the direction of the initial velocity, i.e., $\langle \V{v}(t)|\V{v}_0 \rangle = \frac{\V{v}_0}{|\V{v}_0|} f(t,|\V{v}_0|)$, where $f(t,|\V{v}_0|)$ is a scalar function of time and of the absolute value of the initial velocity.
For sufficiently small initial velocities, this scalar function becomes linearly proportional to $|\V{v}_0|$ and hence one finds
\begin{equation}
\label{VACFrel2}
\langle\V{v}(t)|\V{v}_0\rangle \approx \frac{C(t)}{C(0)}\V{v}_0.
\end{equation}

\subsection{\label{subsec_condvel2}Two Field Variables}

In order to relate the velocity $\V{v}(t)$ of a tagged particle to the velocity field of the fluid, we assume that the velocity of the tagged particle agrees with the velocity of the fluid at its actual position $\V{x}(t)$.
In other words, we assume that $\V{v}(t;\V{v}_0) \stackrel{d}{=}  \mathfrak{u}(\V{x}(t),t)$, where $\mathfrak{u}(\V{x},t)$ is the fluctuating velocity field of the fluid.
Here, $\V{v}(t;\V{v}_0)$ denotes a realization of the velocity of a tagged particle with initial value $\V{v}(0;\V{v}_0) = \V{v}_0$ and $\stackrel{d}{=}$ indicates equality in distribution.    
The conditional velocity average $\langle \V{v}(t)|\V{v}_0 \rangle$ can be expressed in terms of the fluctuating velocity field as
\begin{equation}
\label{assumption}
\langle \V{v}(t)|\V{v}_0 \rangle = \langle \mathfrak{u}(\V{x}(t),t)| \V{v}_0 \rangle, 
\end{equation}
and further transformed as
\begin{equation}
\label{vtv0}
\begin{split}
\langle \V{v}(t)|\V{v}_0 \rangle &= \int d \V{x}\: \langle \mathfrak{u}(\V{x},t) \delta( \V{x} - \V{x}(t))|\V{v}_0\rangle \\
&\approx \int d \V{x}\: \V{u}(\V{x},t;\V{v}_0) n^\mathrm{tag}(\V{x},t;\V{v}_0).
\end{split}
\end{equation}
In going to the second line, we assumed that the velocity field and the tagged particle density are uncorrelated from each other.
Their respective averages are denoted as
\begin{align}
\label{uxt0v0}
\V{u}(\V{x},t;\V{v}_0) &= \langle \mathfrak{u}(\V{x},t)|\V{v}_0 \rangle, \\
\label{ntr0v0}
n^\mathrm{tag}(\V{x},t;\V{v}_0) &= \langle \delta(\V{x} - \V{x}(t)) |\V{v}_0 \rangle.
\end{align}
In order to avoid a too clumsy notation, we suppressed the dependence of the average fields $\V{u}(\V{x},t;\V{v}_0)$ and $n^\mathrm{tag}(\V{x},t;\V{v}_0)$ on the initial position $\V{x}(0) \equiv \V{x}_0$ of the tagged particle.
Without restriction we may choose $\V{x}_0=\V{0}$ by using a proper coordinate system.
Further we note that as a scalar density of a tagged particle in an otherwise isotropic fluid in equilibrium, $n^\mathrm{tag}(\V{x},t;\V{v}_0)$ can only depend on the scalar quantities, $|\V{x}|^2$, $|\V{v}_0|^2$, and $\left(\V{x} \cdot \V{v}_0 \right )^2$.
For sufficiently small velocities $\V{v}_0$, a possible dependence on the last two invariants can be neglected because both are quadratic in the velocity and hence 
\begin{equation}
\label{ntr}
n^\mathrm{tag}(\V{x},t;\V{v}_0) \approx n^\mathrm{tag} (\V{x},t).
\end{equation}
Combining \eqref{VACFrel1}, \eqref{vtv0}, and \eqref{ntr}, one obtains, for the VACF,
\begin{equation}
\label{Ct1}
C(t) \approx \int d \V{v}_0\: \Phi(\V{v}_0)  \V{v}_0 \cdot \int d \V{x}\: \V{u}(\V{x},t;\V{v}_0 ) n^\mathrm{tag}(\V{x},t).
\end{equation} 

For sufficiently small velocities $\V{v}_0$, the absolute value of the velocity field $\V{u}(\V{x},t;\V{v}_0)$ is proportional to the absolute value $v_0 \equiv |\V{v}_0|$.
This allows one to express the velocity field by its average over all absolute values of the initial velocity as
\begin{equation}
\label{avu}
\V{u}(\V{x},t; \V{v}_0) = \frac{v_0}{\overline{v_0}} \overline{\V{u}(\V{x},t;\V{v}_0)},
\end{equation}
where the bar, $\overline{\bullet} \equiv \int dv_0\: \varphi(v_0) \bullet$, indicates a thermal average over the absolute values $v_0$ with respect to the Maxwell--Boltzmann distribution $\varphi(v_0) = m v_0 /(k_\mathrm{B} T) \exp (-m v^2_0/(2 k_\mathrm{B} T))$ \ for $d=2$ with $m$ the mass of the tagged particle, $k_\mathrm{B}$ the Boltzmann constant, and $T$ the temperature.
Note that $\overline{\V{u}(\V{x},t;\V{v}_0)} = \overline{\V{u}(\V{x},t;v_0 \V{e}_{\V{v}_0})}$ only depends on the direction of $\V{v}_0$ denoted by $\V{e}_{\V{v}_0}\equiv \V{v}_0/v_0$ but not on $v_0$.
Therefore, upon replacing the velocity $\V{u}(\V{x},t;\V{v}_0)$ in \eqref{Ct1} by the right-hand side of \eqref{avu}, one can perform the integral over the absolute value of the initial velocity and find
\begin{equation}
\label{Ct2}
C(t) \approx \frac{C(0)}{\overline{v_0}} \int d\V{x}\: \V{e}_{\V{v}_0} \cdot \overline{\V{u}(\V{x},t;v_0 \V{e}_{\V{v}_0})} \: n^\mathrm{tag}(\V{x},t).
\end{equation}   
Here, due to the isotropy of the fluid, $\V{e}_{\V{v}_0} \cdot \overline{\V{u}(\V{x},t;v_0 \V{e}_{\V{v}_0})}$ is independent of the orientation $\V{e}_{\V{v}_0}$.
In Section~\ref{sec_res}, we set $\V{e}_{\V{v}_0}$ to the standard unit vector $\V{e}_x$ along the $x$-axis and compute $\V{u}(\V{x},t) \equiv \overline{\V{u}(\V{x},t;v_0\V{e}_x)}$ as well as $n^\mathrm{tag}(\V{x},t) \equiv \overline{n^\mathrm{tag}(\V{x},t;v_0 \V{e}_x})$ from MD simulations.

\subsection{\label{subsec_hydro}Hydrodynamic Description}

Within the hydrodynamic description, the spreading of the tagged particle density is described by a diffusion equation of the form
\begin{subequations}
\label{PxtPx0}
\begin{align}
\label{Pxt}
\frac{\partial n^\mathrm{tag}(\V{x},t)}{\partial t} &= D(t)\lapl n^\mathrm{tag}(\V{x},t),\\
\label{Px0}
n^\mathrm{tag}(\V{x},0) &= \delta(\V{x}),
\end{align}
\end{subequations}
where, without loss of generality, we choose the origin for the starting point of the tagged particle, i.e., $\V{x}_0 = \V{0}$.
As already mentioned above, in order to properly account for the effects of the long-time tails, we allow for a time-dependent diffusion coefficient, which is directly related to the VACF by $D(t) = \frac{1}{d} \int_0^t C(t') dt'$.
This approach will also allow us to consider the dynamics at short molecular time scales.
Molecular expressions for $D(t)$ and other transport coefficients are presented in Appendix~\ref{appendix_GreenKubo}. 
The solution of the diffusion equation on a square with side length $L$ and periodic boundary conditions can be conveniently given for the Fourier transform of the tagged particle density, $\tilde{n}^\mathrm{tag}(\V{k},t) = \int d\V{x}\: n^\mathrm{tag}(\V{x},t) e^{-i \V{k} \cdot \V{x}}$, where $\V{k} = 2 \pi \V{n}/L$ with $\V{n} = (n_1,n_2,\dots,n_d)$ for $n_1,\dots,n_d=0,\:\pm1,\:\pm2,\dots$.
It becomes
\begin{equation}
\tilde{n}^\mathrm{tag}(\V{k},t) = e^{- k^2 \int_0^t D(t') dt'}.
\end{equation}

According to the hydrodynamic description, the velocity field $\V{u}(\V{x},t;\V{v}_0)$ is given by the solution of the LNS equations (see Appendix~\ref{appendix_LNS}) subject to the initial condition
\begin{equation}
\label{ux0}
\V{u}(\V{x},0;\V{v}_0)=\frac{1}{\bar{n}}\delta(\V{x})\V{v}_0,
\end{equation}
where $\bar{n}$ is the mean number density of the molecular fluid.
As  for the MD simulations, we assumed that the tagged particle exciting the velocity field is initially located at $\V{x}_0 = \V{0}$ and moves with velocity $\V{v}_0$ relative to the resting fluid at this moment.
The resulting solution can be split into the divergence-free component $\V{u}^\perp$ and the rotation-free component $\V{u}^\parallel$ according to the Helmholtz decomposition:
\begin{equation}
\V{u}=\V{u}^\perp+\V{u}^\parallel,\quad\divg\V{u}^\perp=0,\quad\curl\V{u}^\parallel=\V{0}. 
\end{equation}
The rotation-free contribution $\V{u}^\parallel$ can be represented as a superposition of the sound- and density-modes of the LNS (see Appendix~\ref{appendix_LNS}) and hence propagates as an attenuated wave with sound velocity. 
The heat-mode, which is also a longitudinal mode of the LNS, does not contribute because of the uniformity of the temperature field. 
On the other hand, the divergence-free field $\V{u}^\perp$ coincides with the transverse mode of the LNS. 
It spreads only diffusively:
\begin{equation}
\label{uperpxt}
\frac{\partial}{\partial t} \V{u}^\perp(\V{x},t;\V{v}_0) = \nu(t)\lapl\V{u}^\perp(\V{x},t;\V{v}_0).
\end{equation}
The solution evolving from the initial condition~\eqref{ux0} is readily obtained in the $\V{k}$-space as
\begin{equation}
\tilde{\V{u}}^{\perp}(\V{k},t;\V{v}_0)=\frac{1}{\bar{n}}\left( \V{v}_0-\frac{\V{v}_0\cdot \V{k}}{k^2} \V{k}\right)e^{-k^2\int_0^t{\nu(t')dt'}}.
\label{uperxkt}
\end{equation}
Because the divergence-free part spreads diffusively, it decays much slower than the rotation-free part, which, as just mentioned, propagates with the sound velocity.
Hence, the latter can be neglected as a contribution to the expression~\eqref{Ct1} for the VACF at large times.~\cite{PomeauResibois1975, HansenMcDonald2013}
Finally, using Parseval's theorem, one may express the spatial integral on the right-hand side of \eqref{Ct1} by means of a sum over all $\V{k}$ values, which can be approximated by an integral in the limit of large systems. One then obtains the expression
\begin{equation}
\label{Ctlongtime}
C(t) \approx \frac{(d-1)k_\mathrm{B}T}{\bar{n} m}\frac{1}{\big[4\pi \int_0^t{\left(\nu(t')+D(t')\right)dt'}\big]^{d/2}}.
\end{equation}
To summarize, this result is based on the following assumptions: validity of the LNS equations with time-dependent transport coefficients and also large times and large system size.
Note that, for time-independent transport coefficients $D$ and $\nu$, the well-known algebraic decay expression for the VACF~\cite{ErnstHaugeLeeuwen1971b} is recovered.

\section{\label{sec_num_proc}Model and Numerical Procedure}

\subsection{\label{subsec_sys_MD}System and MD Simulation}

We consider a standard model of a two-dimensional molecular fluid.
It consists of $N$ identical fluid particles with mass $m$ in a square of side length $L$ with periodic boundary conditions.
The fluid particles interact pairwise via a potential function $V(r)$ of the  inter-particle distance $r$.
Hence, the Hamiltonian of the system is given as
\begin{equation}
\label{Hamiltonian}
H=\sum_i\frac{\V{p}_i^2}{2m}+\sum_{i>j}V\left(\left|\V{x}_i-\V{x}_j\right|\right),
\end{equation}
where $\V{x}_i$ and $\V{p}_i=m\V{v}_i$ are the position and momentum of the $i$th fluid particle.
For the pair potential $V(r)$, we employ the Weeks--Chandler--Andersen potential, which is a purely repulsive potential of Lennard-Jones type: 
\begin{equation}
\label{WCApot}
V(r)=
\begin{cases}
4\varepsilon\left[\left(\frac{\sigma}{r}\right)^{12}-\left(\frac{\sigma}{r}\right)^6\right]+\varepsilon & \mbox{for $r \leq 2^{1/6}\sigma$}, \\
0 & \mbox{for $r \geq 2^{1/6}\sigma$}.
\end{cases}
\end{equation}
Here, $\sigma$ is the diameter of a fluid particle and $\varepsilon$ is the interaction strength parameter.
Based on the molecular parameters in \eqref{Hamiltonian} and \eqref{WCApot}, we use reduced (dimensionless) MD units. 
That is, the units of mass, length, and energy are set to $m$, $\sigma$, and $\varepsilon$, respectively, and the units of any other quantities are determined from them (e.g., $\sigma\sqrt{m/\varepsilon}$ for time and $\varepsilon/k_\mathrm{B}$ for temperature).

We simulate a fluid having number density $\bar{n}=0.6$ and temperature $\bar{T}=1$.
This state is chosen so that the algebraic decay of the VACF can be readily observed.
While this long-time behavior is universal, the time scale on which it is clearly seen depends largely on the density of the fluid.~\cite{McDonoughRussoSnook2001, DibOuld-KaddorLevesque2006}
At lower densities, the algebraic decay emerges at later times.
At higher densities, on the other hand, the VACF displays negative values at short times (due to backscattering effects) before the algebraic decay pattern develops.
Moreover, since it takes less time for sound waves to travel across the domain, the disturbance of the long-time tail occurs at earlier times. 
To identify finite size effects, we simulate three system sizes having $N=512$, 1024, and 2048 fluid particles, which correspond to domain sizes $L=\sqrt{N/\bar{n}}=29.2$, 41.3, and 58.4, respectively.
The mean free path and the mean collision time are roughly estimated to be 0.2 and 0.2, respectively, from a corresponding hard-disk system having the same number density and temperature.~\cite{GaspardLutsko2004}

We perform $NVE$ simulations~\cite{Rapaport2004} using the standard velocity Verlet algorithm with timestep size $\D{t}_\mathrm{MD}=10^{-3}$.
Equilibrium samples are prepared as follows.
For each initial configuration where the positions and momenta of the fluid particles are randomly chosen with zero total momentum, velocity scaling with the target temperature $\bar{T}=1$ is performed 10 times every $10^3$ steps and then equilibration is performed for additional $10^5$ steps.
While performing a single run for these simulations is not computationally expensive at all, we generate a large $NVE$-ensemble of $\mathcal{N} =1024$ MD trajectories in order to obtain smooth flow patterns.

\subsection{\label{subsec_avg_proc}Averaging Procedure for Field Quantities}

The number density $n(\V{x},t)$ and the velocity $\V{u}(\V{x},t)$ of a fluid, generated by the motion of a tagged particle, together with the resulting tagged particle density $n^\mathrm{tag}(\V{x},t)$ are expressed as averages over MD trajectories of the individual fluid particles in terms of the formal expressions  
\begin{subequations}
\label{def_field}
\begin{align}
\label{num_dens}
&n(\V{x},t) = \Big\langle\!\!\Big\langle \sum_i \delta \left( \V{x}_i(t) - \V{x} \right) \Big\rangle\!\!\Big\rangle, \\
\label{vel}
&\V{u}(\V{x},t) n(\V{x},t) = \Big\langle\!\!\Big\langle \sum_i \V{v}_i(t) \delta \left( \V{x}_i(t) - \V{x} \right)\Big\rangle\!\!\Big\rangle, \\
\label{num_dens_tr}
&n^\mathrm{tag}(\V{x},t) = \Big\langle\!\!\Big\langle \delta \left( \V{x}_1(t) - \V{x} \right) \Big\rangle\!\!\Big\rangle.
\end{align}
\end{subequations}
Here, the particle~1 is considered as the tagged particle. 
The double brackets $\langle\!\langle\;\rangle\!\rangle$ denote the thermal equilibrium average conditioned on the specific initial position and initial velocity-direction of the tagged particle at $\V{x}_1(0)=\V{0}$ and $\V{v}_1(0)/|\V{v}_1(0)| = \V{e}_x$, respectively. 

In order to comply with this initial condition in those cases when the actual initial data of the particle~1 deviate from the required values, we used the homogeneity of the configuration space and also assumed its homogeneity by applying to each particle the translation $\V{T}$ that shifts the particle~1 to the origin ($\V{T} \V{x}_1(0) =\V{0}$), and a rotation $\V{R}$ that brings its velocity into the positive $x$-direction ($\V{R}\V{v}_1(0) = |\V{v}_1(0)| \V{e}_x$) yielding transformed positions ($\hat{\V{x}}_i(t) =\V{R} \V{T}\V{x}_i(t)$) and velocities ($\hat{\V{v}}_i(t) = \V{R} \V{v}_i(t)$) of all particles.
Strictly speaking, the configuration space underlying the MD simulations, which is a two-dimensional torus due to the periodic boundary conditions, does not have rotations as symmetry operations, because they fail to be bijective.
However, as long as the flow pattern has not covered the full torus but is rather concentrated in a region around the origin, one may rotate this pattern as if it were defined on the entire Euclidean plane. 

Due to the permutation symmetry with respect to the numbering of particles, any other particle~$j$ can be also assigned as the tagged particle and the resulting set of trajectories can be used to perform the averages in \eqref{def_field}.
Moreover, the time-translational invariance of the trajectories in thermal equilibrium allows one to take any time point $l\D{t}$ as initial time, yielding new trajectories
\begin{equation}
\label{traj}
\hat{\V{x}}^\paren{j,l}_i(k\D{t}) = \T{R}^\paren{j,l}\T{T}^\paren{j,l}\V{x}_i((k+l)\D{t}), \quad \hat{\V{v}}^\paren{j,l}_i(k\D{t}) = \T{R}^\paren{j,l}\V{v}_i((k+l)\D{t}),
\end{equation}
where $k=0, \ldots, N_2$ labels the $l$th trajectory with $l = 0, \ldots, N_1-N_2$ and $N_2 \ll N_1$.
Here, $\D{t}$ is an integer multiple of the timestep $\D{t}_\mathrm{MD}$ and the translation $\V{T}^{(j,l)}$ and rotation $\V{R}^{(j,l)}$ are such that the initial tagged particle position $\V{x}_j(l\D{t})$ is shifted to the origin and its velocity $\V{v}_j(l\D{t})$ is rotated into the positive $x$-direction.
Hence, the brackets in \eqref{def_field} denote an average over a large ensemble generated by $\mathcal{N}$ MD runs with all particles at any time generating an initial value of trajectories as specified in \eqref{traj}. 

In order to estimate the averaged fields~\eqref{def_field}, we discretized the state space by introducing square cells of side length $\D{x}$.
The Dirac delta functions were approximated by means of the indicator function   $\mathbb{I}(\mathcal{C})$, which is $1$ if the condition $\mathcal{C}$ is true and $0$ otherwise.
The three fields in \eqref{def_field} were estimated in the following way:
\begin{subequations}
\label{fieldest}
\begin{align}
n(\V{x},k\D{t}) &\approx \frac{1}{N(N_1\!-\!N_2\!+\!1)}\Big \langle \sum_\paren{j,l}\sum_i \frac{1}{\D{x}^2}\mathbb{I}\left[\hat{\V{x}}^\paren{j,l}_i(k\D{t})\in \mbox{cell $\V{x}$}\right] \Big \rangle_{\mathcal{N}}\:,\\
\V{u}(\V{x},k\D{t})n(\V{x},k\D{t}) &\approx \frac{1}{N(N_1\!-\!N_2\!+\!1)} \nonumber\\
&\quad \times \Big \langle \sum_\paren{j,l}\sum_i \hat{\V{v}}_i^\paren{j,l}(k\D{t}) \frac{1}{\D{x}^2}\mathbb{I}\left[\hat{\V{x}}^\paren{j,l}_i(k\D{t})\in \mbox{cell $\V{x}$}\right] \Big \rangle_{\mathcal{N}}\:,\\
n^\mathrm{tag}(\V{x},k\D{t}) &\approx \frac{1}{N(N_1\!-\!N_2\!+\!1)}\Big \langle \sum_\paren{j,l} \frac{1}{\D{x}^2}\mathbb{I}\left[\hat{\V{x}}^\paren{j,l}_j(k\D{t})\in \mbox{cell $\V{x}$}\right] \Big \rangle_{\mathcal{N}} \:,
\end{align}
\end{subequations}
where $\langle \bullet \rangle_{\mathcal{N}}$ denotes the arithmetic average over the results of $\mathcal{N}$ independent MD runs.            
The average fields were determined using $\D{x}=0.1$, $\D{t}=500 \D{t}_{\text{MD}}=0.5$, $N_1=4020$ (for $N=512$ and 1024), 2020 (for $N=2048$), and $N_2=20$.
Thus, the field quantities were computed up to the time $N_2\D{t}=10$ from MD trajectories of length $N_1\D{t}\approx 2\times 10^6 \D{t}_\mathrm{MD}$ (for $N=512$ and 1024) and $10^6\D{t}_\mathrm{MD}$ (for $N=2048$).
A total of $\mathcal{N}=1024$ MD samples were generated.

\subsection{Helmholtz Decomposition}

We determined the longitudinal and transversal components of the velocity field $\V{u}(\V{x},t)$ obtained from the MD simulations using the Helmholtz decomposition.
In the two-dimensional case, this decomposition is obtained from two scalar fields $\Phi(\V{x},t)$ and $A(\V{x},t)$ as~\cite{BhatiaNorgardPascucciBremer2014}
\begin{equation}
\label{helmholtzdecomp} 
\V{u}(\V{x},t)=-\grad\Phi(\V{x},t)+\T{J}\grad A(\V{x},t),
\end{equation}
where $-\grad\Phi(\V{x},t)$ is the rotation-free component corresponding to the longitudinal velocity $\V{u}^{\parallel}(\V{x},t)$ and $\T{J}\grad A(\V{x},t)$ is the divergence-free component corresponding to the transverse velocity $\V{u}^{\perp}(\V{x},t)$.
Here, $\T{J}$ denotes the counterclockwise rotation by $\pi/2$.

Both scalar fields are solutions of the Poisson equation. The source terms of $\Phi(\V{x},t)$ and $A(\V{x},t)$ are determined by the divergence and the rotation of the velocity field, respectively, i.e.,
\begin{subequations}
\begin{align}
\label{Poissonlongi}
&\lapl\Phi(\V{x},t) = -\divg \V{u}(\V{x},t),  \quad \mbox{for $\V{x}\in\Omega$},\\
\label{Neumannlongi}
&\V{n}\cdot\grad\Phi(\V{x},t) = -\V{n}\cdot\V{u}(\V{x},t) \quad \mbox{for $\V{x}\in\partial\Omega$},
\end{align}
\end{subequations}
and
\begin{subequations}
\begin{align}
\label{Poissontrans}
&\lapl A(\V{x},t) = -\divg \V{J} \V{u}(\V{x},t),\quad \mbox{for $\V{x}\in\Omega$},\\
\label{Neumanntrans}
&\V{n}\cdot\grad A(\V{x},t) = -\V{n}\cdot\V{J}\V{u}(\V{x},t) \quad \mbox{for $\V{x}\in\partial\Omega$},
\end{align}
\end{subequations}
where $\V{n}$ is the outward normal to the boundary.
Because the velocity field does not strictly obey periodic boundary conditions 
due to the construction of the ensemble on which \eqref{fieldest} is based, we used Neumann boundary conditions.

\subsection{Comparison with LNS}

Below we compare the velocity fields and their longitudinal as well as transversal components obtained from the MD simulations with solutions of the LNS equations, which are presented in Appendix~\ref{appendix_LNS}. 
The initial condition for the velocity field complies with \eqref{ux0}, whereas the fluid density and temperature fields initially are assumed as uniform:
\begin{equation}
\label{nx0Tx0}
n(\V{x},0)=\bar{n},\quad T(\V{x},0)=\bar{T},
\end{equation}
where $\bar{n}$ and $\bar{T}$ are the mean number density and temperature of the system, respectively.

All parameters needed to solve 
the LNS equations are determined by the MD simulations.
As already mentioned, the time-dependent transport coefficients entering the LNS equations are obtained from Green--Kubo-like expressions that are presented in Appendix~\ref{appendix_GreenKubo}.
Thermodynamic variables, such as the adiabatic speed of sound $c_\mathrm{s}$, the ratio of specific heats $\gamma$, and the thermal expansion coefficient $\alpha$ are computed by the method of pressure derivatives,~\cite{MeierKabelac2006} see also Appendix~\ref{appendix_thermo}.
As numerical values for the present two-dimensional system, we obtained $c_\mathrm{s}=4.43$, $\gamma=1.83$, and $\alpha=0.31$.

\section{\label{sec_res}Results and Discussion}

In Section~\ref{subsec_qual_obs}, we give a brief description of the field variables that were computed from the MD simulations. 
In Section~\ref{subsec_comp_lns}, we compare these results with the hydrodynamic description.
In Section~\ref{subsec_vacf}, we examine the validity of the assumptions implied by the hydrodynamic description of the long-time VACF and discuss the applicability and limitations of the description.

\subsection{\label{subsec_qual_obs}Field Variables}

\begin{figure}
\centerfloat
\includegraphics[width=0.9\linewidth]{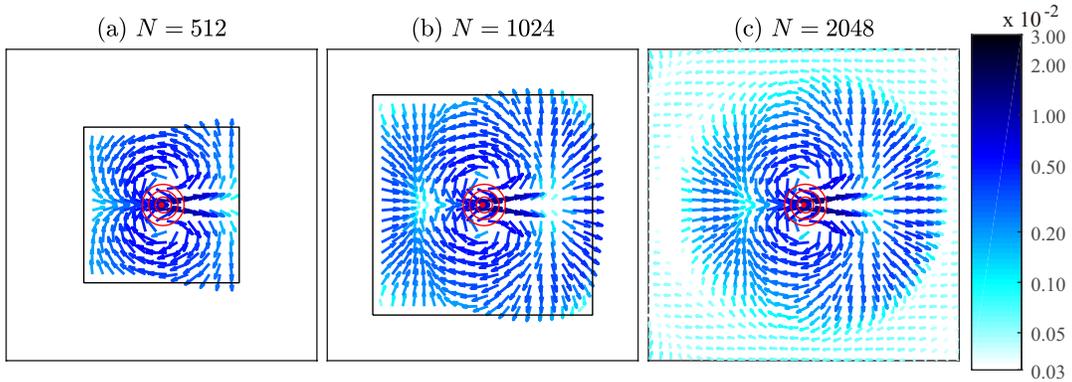}
\caption{\label{fig_flow}
Velocity field $\V{u}(\V{x},t)$ at $t=4$.
The flow patterns obtained from MD simulations are presented in panels~(a)--(c) in an increasing order of system size, $N=512$, 1024, and 2048 ($L=29.2$, 41.3, and 58.4). 
The entire domain is divided into cells of side length $\D{x}_\mathrm{vel}=2$ and the average velocity of each cell is depicted by an arrow, which is colored depending on the log scale of its magnitude.
The red dot at the center denotes the initial position of the tagged particle.
The three contours with red solid lines depict regions within which the tagged particle is found at probabilities 0.5 (inner), 0.9 (middle), and 0.99 (outer).
The vector fields for the two small system sizes clearly exhibit visible deviations from periodicity at the boundaries.
Only for the largest system, the flow field generated by the tagged particle does not yet cover the full square and hence is not influenced by the finite system size at time $t=4$.}
\end{figure}

\begin{figure}
\centerfloat
\includegraphics[width=0.9\textwidth]{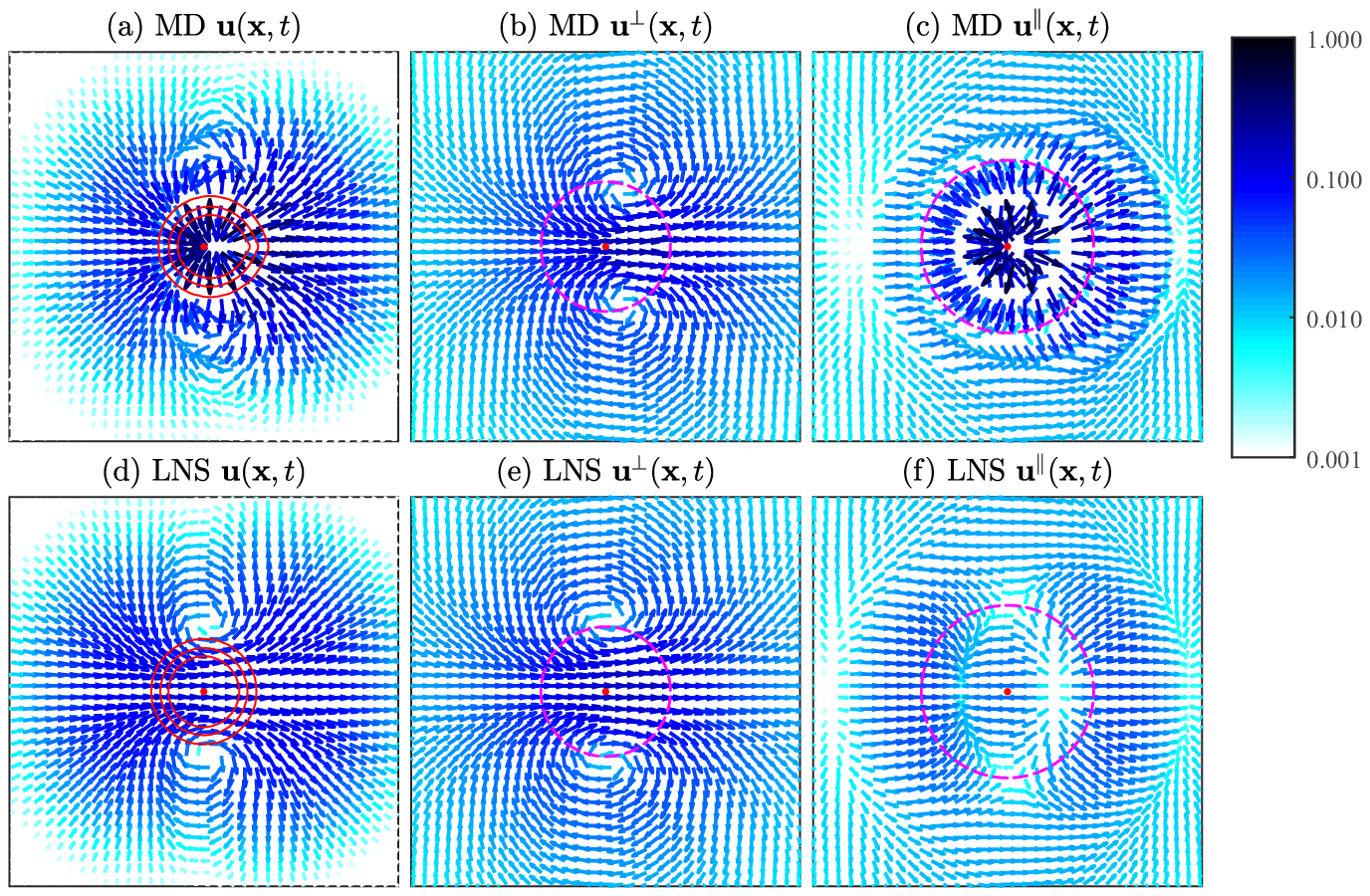}
\caption{\label{t05}
Velocity field $\V{u}(\V{x},t)$ at $t=0.5$ and its Helmholtz decomposition.
The velocity fields resulting from the MD simulations for a system of $N=2048$ particles (upper row) and from the solution of the LNS equations (bottom row) are displayed in the left column.  
The middle and the right columns present the perpendicular and the parallel components, respectively.
Each panel displays the central square $(-5,5)\times(-5,5)$ of the configuration space.
The average velocity of each square cell of side length $\D{x}_\mathrm{vel}=0.3$ is depicted by an arrow.
The magnitude of the velocities is indicated by the log-scale length of the corresponding arrow and a color-code.
At this relatively short time, the velocity field based on the MD simulations is still rather strong near the center as seen in panel~(a), whereby the large contributions result from the parallel field presented in panel~(c). 
The perpendicular field component exhibits a counter-rotating vortex pair in panel~(b). 
This flow pattern is quite well reproduced by the transverse part of the solution of the LNS equations as evidenced by panel~(e). 
A large discrepancy is found for the parallel velocity fields, see panels~(c) and (f), which is also reflected in the total velocity fields in panels~(a) and (d). 
The three contours in panels~(a) and (d) border the regions within which one finds the tagged particle with probabilities $0.5$ (inner), $0.9$ (middle), and $0.99$ (outer) according to MD and LNS, respectively.
The magenta circles in panels~(b) and (e) indicate how far the centers of the vortices have moved according to \eqref{vortexy0}.
Finally, in panels~(c) and (f), the magenta circles characterize the propagation of a signal with sound velocity $c_\mathrm{s}$ initially emitted from the center.}
\end{figure}

\begin{figure}
\centerfloat
\includegraphics[width=0.9\textwidth]{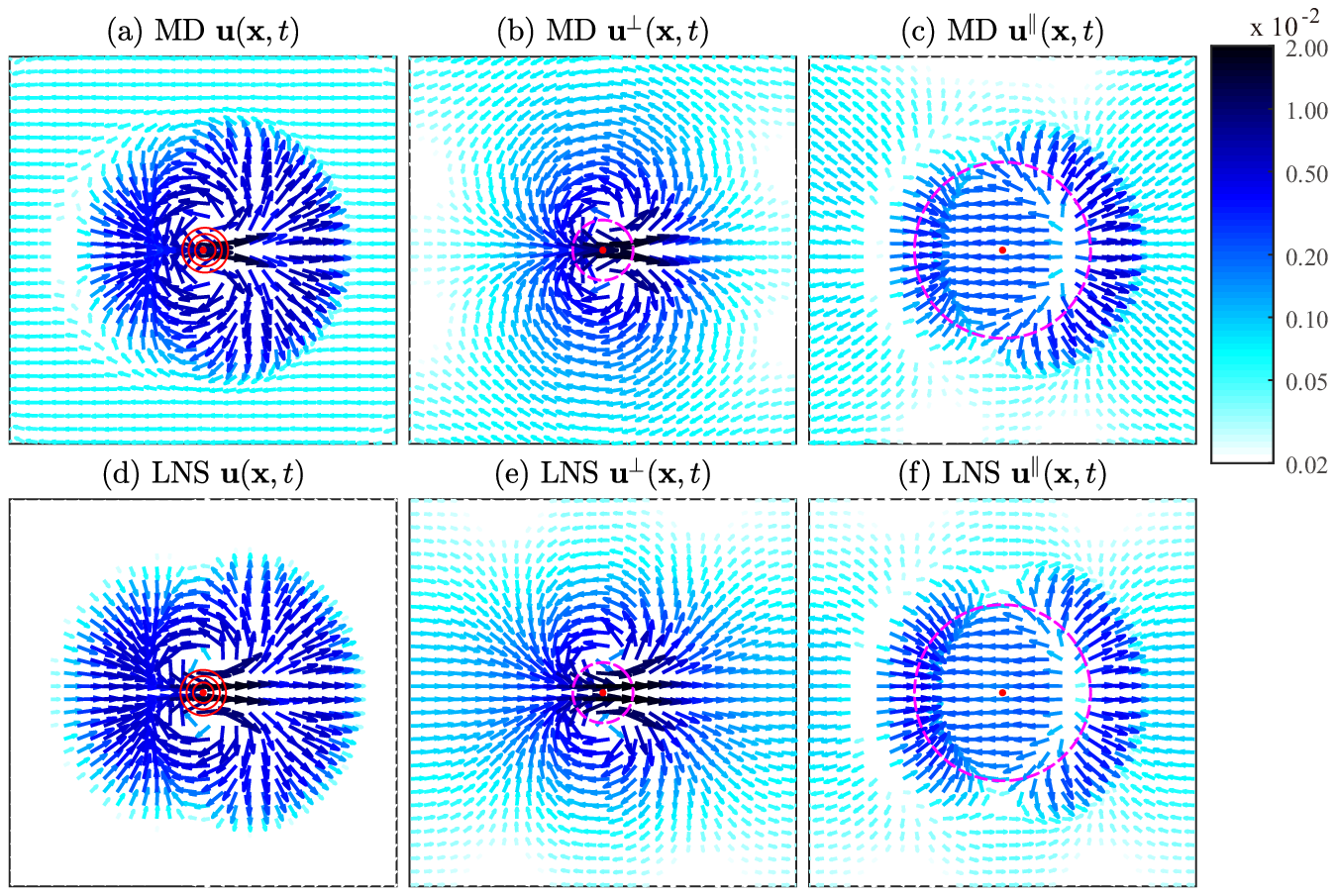}
\caption{\label{t3}
Velocity field $\V{u}(\V{x},t)$ at $t=3$ and its Helmholtz decomposition.
The velocity fields resulting from the MD simulations for a system of $N=2048$ particles (upper row) and from the solution of the LNS equations (bottom row) are displayed in the left column. 
The middle and the right columns present the perpendicular and the parallel components, respectively.   
Each panel displays the domain $(-L/2,L/2)\times(-L/2,L/2)$ with $\D{x}_\mathrm{vel}=2$.
At this still relatively short time $t=3$, which roughly corresponds to 15 mean collision times between fluid particles, the agreement between MD and LNS is surprisingly good. 
The strongest deviations exist for the parallel fields, see panels~(c) and (f). 
In particular, the front on the right-hand side of the center is more pronounced for the MD results than for the LNS. 
The magenta circles in panels~(b) and (e) indicate how far the centers of the two vortices have diffusively propagated up to time $t=3$ according to \eqref{vortexy0}. 
The much larger circles in panels~(c) and (f) specify the distance covered by a sound wave propagating at speed $c_\mathrm{s}=4.43$.  
Note that the velocity fields obtained from the MD simulations still satisfy quite well periodic boundary conditions while the LNS fields are exactly periodic.}
\end{figure}

The velocity fields $\V{u}(\V{x},t)$, at time $t=4$, obtained from the MD simulations for three system sizes ($N=512$, 1024, 2048), are displayed in Fig.~\ref{fig_flow}.
In all cases the velocity field is caused by a tagged particle initially moving in the positive $x$-direction.
The magnitude of the field is obtained from an average over all absolute values of the initial velocity of the tagged particle according to the Maxwell--Boltzmann distribution as introduced in \eqref{avu}.
For the small and middle sized systems the resulting flow patterns cover the entire square, while the flow field of the largest system is still localized around the initial position of the tagged particle.
The distribution of the tagged particle $n^\mathrm{tag}(\V{x},t)$ is indicated by three contours within which the particle is found with probabilities 0.5, 0.9, and 0.99 when going from the inner most to the outer circle.
This distribution is well described by a Gaussian, which is slightly shifted to the right due to the directional bias of the initial velocity of the tagged particle, see also Fig.~\ref{fig_ntr}. 
As expected from the fact that the mass diffusion is much slower than the momentum transport in a dense fluid, $n^\mathrm{tag}(\V{x},t)$ spreads slower than $\V{u}(\V{x},t)$. 
The fluid density $n(\V{x},t)$ (not shown) has a background value $\bar{n}$ with an atomistic structure which has the same origin as the peaks found in the pair correlation function.

Whereas the velocity field $\V{u}(\V{x},t)$ contains both vorticity and sound-wave contributions, a clear separation into $\V{u}^\perp(\V{x},t)$ and $\V{u}^\parallel(\V{x},t)$ is achieved by the Helmholtz decomposition as presented in the upper rows of Fig.~\ref{t05} (at $t=0.5$) and Fig.~\ref{t3} (at $t=3$).
The sizes of the patterns represented by the perpendicular and the longitudinal fields clearly reflect the fast wave-like expansion of the latter and the slow diffusional motion of the former.
Figure~\ref{t05} exemplifies the flow patterns at the time $t=0.5$ corresponding to the two- to three-fold of the mean collision time.
At this very short time, the velocity field has still very large values close to the center.
Yet panel (b) displays a vortex pair that has already developed above and below  the center.
The largest velocity contributions to $\V{u}(\V{x},0.5)$ though come from the parallel component presented in panel (c).
At the somewhat later time $t=3$, these large velocity contributions have been already dissipated leaving a vortex and a wave-like contribution as exemplified by Fig.~\ref{t3}~(a)--(c).   

When the sound wave has propagated by the distance $L/2$, the flow pattern obtained from rotated MD configurations starts failing to obey the periodic boundary conditions.
The upper limit of time before this happens is given by  $t_{\mathrm{max}}=L/(2c_{\mathrm{s}})$, which corresponds to 3.23, 4.66, and 6.60 for $N=512$, 1024, and 2048, respectively.
The finite size effect becomes visible at larger times. 
Due to the diffusive character, the shear mode propagation and the mass diffusion are slower than the sound wave propagation and finite size effects on $\V{u}^\perp(\V{x},t)$ and $n^\mathrm{tag}(\V{x},t)$ only appear at  later times.

Even though the identification of the vortex pairs in panels~(b) of Figs.~\ref{fig_flow}--\ref{t3} appears straightforward from an intuitive point of view, we corroborated their existence by two objective criteria, see Appendix~\ref{appendix_Gal}.

\subsection{\label{subsec_comp_lns}MD versus LNS}

In the second rows of Figs.~\ref{t05} and \ref{t3}, the solutions of the LNS equations~\eqref{LNS}, together with the corresponding transverse and longitudinal components, are displayed for $t=0.5$ and $t=3$.
While at the very short time $t=0.5$ large deviations between the MD (depicted in the upper row) and the LNS velocity fields exist, the transverse components (displayed in the middle column) already agree surprisingly well.
After the still rather short time $t=3$, the agreement between the MD and the LNS velocity fields is very good.
The largest deviations can be seen in the parallel field components.
The deviations between MD and LNS have two distinct origins.
One reason is the continuum nature of the Navier--Stokes equations, which is in manifest contrast to the atomistic structure underlying the MD simulations.
An additional reason for the observed differences is the approximation of the nonlinear Navier--Stokes equations by the linearized equations~\eqref{LNS}, which is not justified for the considered singular initial condition at least at short times. 

\begin{figure}
\centerfloat
\includegraphics[width=\linewidth]{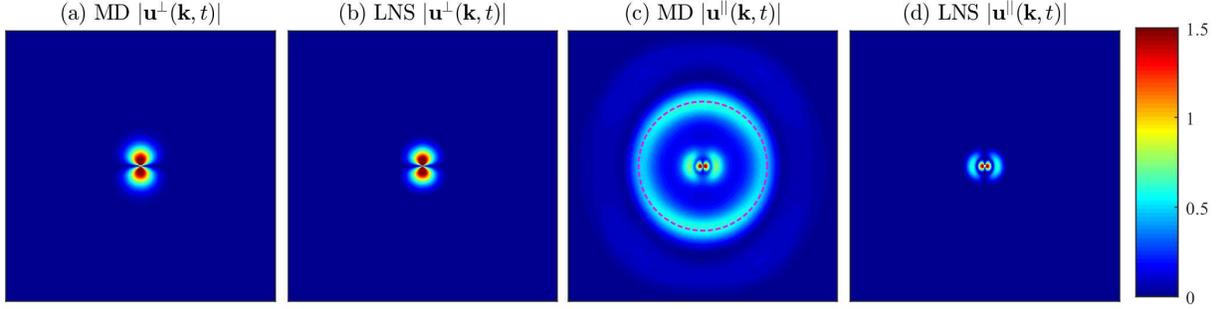}
\caption{\label{fig_uk}
Comparison in the $\V{k}$-space at $t=0.5$.
For the longitudinal ($\V{u}^\parallel$) and transverse ($\V{u}^\perp$) velocity components, the MD results are compared with the LNS solutions \eqref{ukt} by the absolute values of Fourier modes $|\tilde{\V{u}}^{\bullet}(\V{k},t)|$ for $\bullet =\perp,\parallel$ and $\V{k}\in(-12,12)\times(-12,12)$. 
In panel~(c), a circle of radius $2\pi/r_1$ is drawn by the magenta dashed line for comparison, where $r_1$ is the mean nearest-neighbor distance.
MD results from the largest system $N=2048$ are used.}
\end{figure}

In order to get a better understanding to which extent the atomistic structure of the fluid plays a role, we pass from the $\V{x}$-space to the $\V{k}$-space and compare magnitudes of the Fourier transformed fields  $\tilde{\V{u}}^{\bullet}(\V{k},t) = \int d\V{x}\:\V{u}^{\bullet}(\V{x},t) e^{-i \V{k} \cdot \V{x}}$ for $\bullet =\perp,\parallel$ resulting from the MD simulations with the corresponding LNS fields, which are explicitly given by \eqref{ukt}.        
In Fig.~\ref{fig_uk}, the absolute values $|\tilde{\V{u}}^{\bullet}(\V{k},t)| = \left( |\tilde{u}_x^\bullet|^2+|\tilde{u}_y^\bullet|^2 \right)^{1/2}$, $\bullet = \perp, \parallel$ at time $t=0.5$ are compared.
While, as in direct space, the agreement between the MD and LNS results for the perpendicular contribution is almost perfect, the MD result for the parallel component displays distinct structures at large $k$ values, while the differences at small $k$ values are minor. 
The most pronounced difference is a ridge along a circle with the radius $k = 2 \pi/r_1$ corresponding to the nearest neighbor distance $r_1=1.096$ in the fluid.
The appearance of a maximum at this particular $k$ value clearly reflects the influence of the atomistic structure of the fluid on the MD velocity field.
Beyond this radius the absolute velocity $|\tilde{\V{u}}^\parallel(\V{k},0.5)|$ first decreases then develops a broad and shallow maximum at even larger $k$ values.
The amount of quasi-momentum transferred to small $k$ values is restricted to two regions around $2 \pi n/L \V{e}_{\V{x}}$ with $n =\pm 1$ according to both MD and LNS methods.
Deviations between the MD and LNS results already at slightly larger $k$ values are clearly visible.
Whether they are caused by the atomistic structure of the fluid or by the linearization of the Navier--Stokes equations is not clear.    

\begin{figure}
\centerfloat
\includegraphics[width=\linewidth]{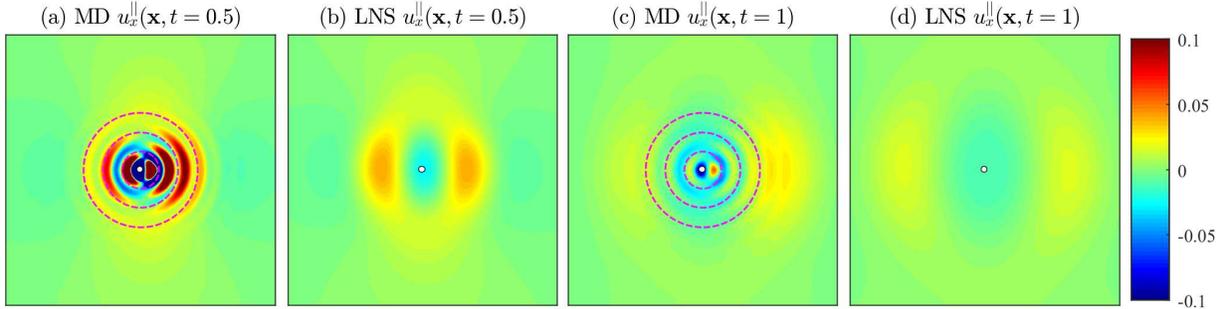}
\caption{\label{fig_upar}
Atomistic structures observed only in the MD results.
For the $x$-component of the longitudinal velocity $u^\parallel_x(\V{x},t)$, the MD results at $t=0.5$ and 1 are compared with the inverse Fourier transformed results of LNS solution~\eqref{ukt} for $\V{x}\in(-8,8)\times(-8,8)$.
In panels~(a) and (c), concentric circles of radii $r_i$ ($i=1,2,3$) centered at the origin are drawn by the magenta dashed lines for comparison, where $r_i$ is the position of $i$th peak of the radial distribution function.
The white dot at the center denotes the initial position of the tagged particle.
MD results from the largest system $N=2048$ are used.
The LNS results are obtained from the $\V{k}$-sum over $2\pi/L\le |k_x|,|k_y|\le \pi/\D{x}_\mathrm{c}$ with $\D{x}_\mathrm{c}=0.1$.}
\end{figure}

According atomistic structures are also visible in direct space at sufficiently small scales.
Figure~\ref{fig_upar} presents the $x$-component of $\V{u}^\parallel(\V{x},t)$ in the approximately threefold magnified central region at two times $t=0.5$ and $t=1$.
At the smaller time, the MD result displays a shell-like structure, which is determined by concentric circles with radii $r_i$ coinciding with the locations of the first three maxima of the radial distribution function at $r_1=1.096$, $r_2=2.226$, and $r_3=3.390$.
In the central circle with radius $r_1$ and also in the two rings bordered by neighboring radii $r_i$, there are roughly croissant-shaped regions of forward motion surrounded by regions with backward motion. 
In total, the amount of backward motion is less pronounced in accordance with the fact that it is generated by backscattering events.  
These alternating structures of forward and backward motion are much less pronounced at the later time $t=1$, when the initial excitation of the fluid by the tagged particle has already propagated away from the center.
Of course, none of the atomistic structures are present in the LNS results. Also, the LNS velocity patterns displayed in Fig.~\ref{fig_upar} appear to be more symmetric with respect to the $y$-axis than the corresponding MD result.

\begin{figure}
\centerfloat
\includegraphics[width=0.55\linewidth]{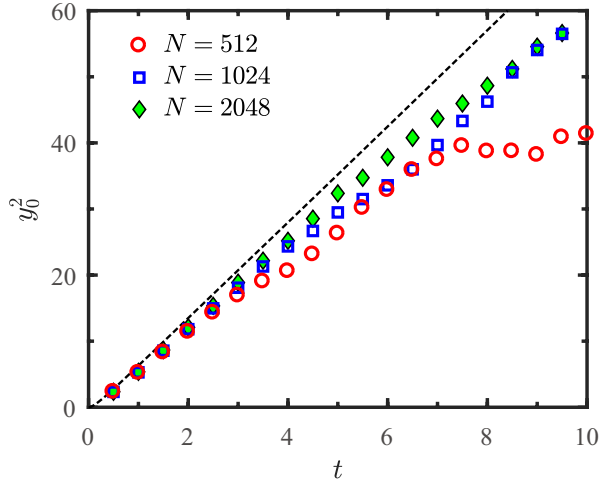}
\caption{\label{fig_y0sq}
Shear mode propagation.
For the vortex centers $(x_0,\pm y_0)$ of the transverse velocity field, the growth of $y_0^2$ is shown versus time $t$.
The circle, square, and diamond symbols denote MD results from the three system sizes $N=512$, 1024, and 2048, respectively. 
The dashed line depicts theoretical prediction~\eqref{vortexy0} from the LNS solution.}
\end{figure}

\begin{figure}
\centerfloat
\includegraphics[width=0.55\linewidth]{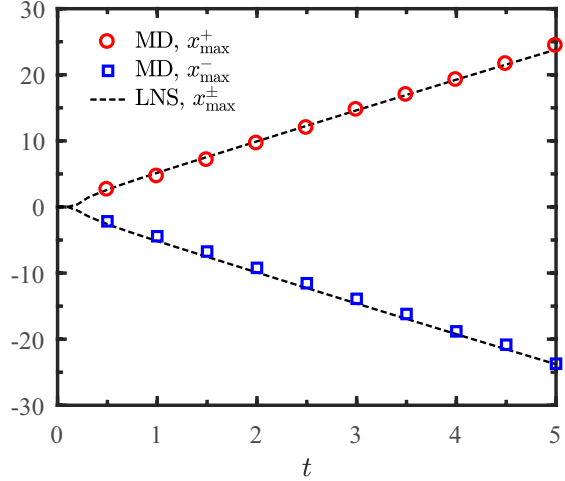}
\caption{\label{fig_cs}
Sound wave propagation.
For the points $(x_\mathrm{max}^+,0)$ and $(x_\mathrm{max}^-,0)$ attaining the maximum magnitude of the longitudinal velocity in the forward and backward wavefronts, respectively, the growth of $x_\mathrm{max}^\pm$ is shown versus time $t$.
The MD results from the largest system size $N=2048$ are depicted by circles ($x_\mathrm{max}^+$) and squares ($x_\mathrm{max}^-$), whereas the LNS predictions $x_\mathrm{max}^+ = -x_\mathrm{max}^-$ are plotted by the dashed lines.}
\end{figure}

According to the LNS equations, the centers of the two vortices, defined as the locations where $\V{u}(\V{x},t)$ vanishes, move in the directions perpendicular to the initial tagged particle velocity in proportion to $[\int \nu(t') dt']^{1/2}$, see \eqref{vortexy0}. 
In Fig.~\ref{fig_y0sq}, this prediction is compared with the movement of the respective locations of the perpendicular velocity component determined by MD simulations. The agreement is good as long as finite size effects can be neglected.
The LNS motion of the vortices is strictly confined to the $y$ direction in contrast to the results obtained from the MD simulation according to which the vortices also move into the $x$-direction approaching the maximal distance $x_0\approx 0.6$.
Apart from this minor difference, the agreement of the vortex pattern found in the MD simulations and obtained from LNS is excellent. 
Likewise we compared the speed of the wave-like propagation in Fig.~\ref{fig_cs} and found that both front positions $x^{\pm}_{\text{max}}$, where $|\V{u}^\parallel(\V{x},t)|$ attains maxima on the $x$-axis, move according to the MD simulations with the sound speed in complete agreement with the LNS. 

\begin{figure}
\centerfloat
\includegraphics[width=1.1\linewidth]{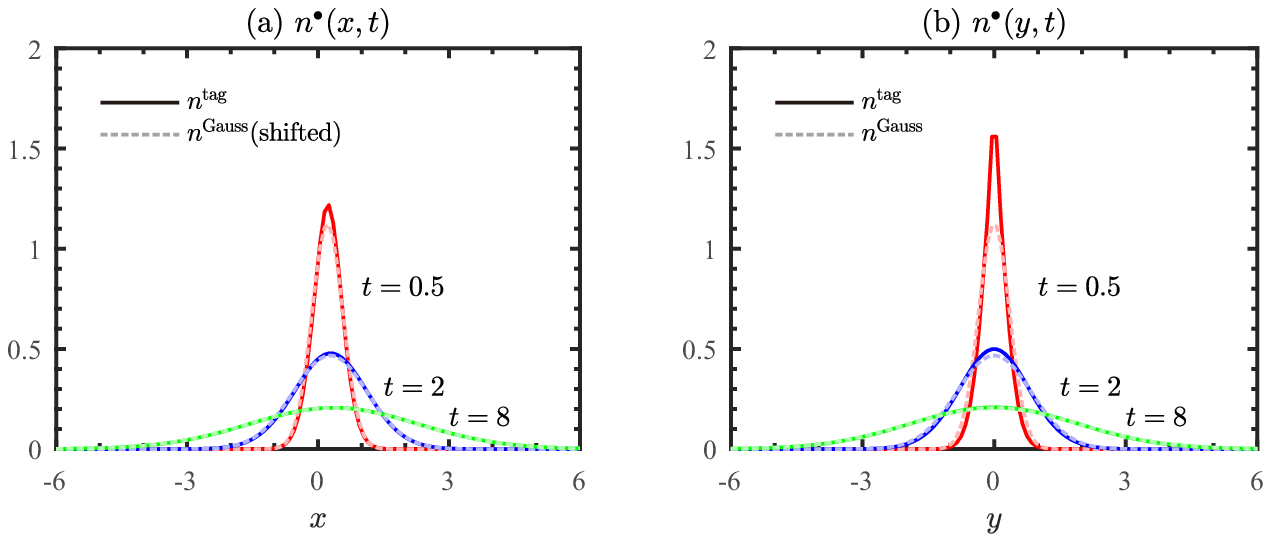}
\caption{\label{fig_ntr}
Distribution of the tagged particle.
The marginal densities $n^\mathrm{tag}(x,t)\equiv\int n^\mathrm{tag}(x,y,t) dy$ and $n^\mathrm{tag}(y,t)\equiv\int n^\mathrm{tag}(x,y,t) dx$ of the tagged particle resulting from MD simulations (solid lines) are compared with the Gaussian densities according to \eqref{PxtPx0} (dashed lines), $n^\mathrm{Gauss}(x-\langle x(t)\rangle,t)$ (shifted) and $n^{\text{Gauss}}(y,t)$, respectively, in panels (a) and (b).
Results from a system of size $N=2048$ are shown for different values of time $t=0.5$ (red), 2 (blue), and 8 (green).
The agreement is excellent for the two larger times $t=2$ and $t=8$.
For the short time $t=0.5$, the MD result is narrower than the Gaussian distribution due to a dependence of $n^\mathrm{tag}(x,y,t)$ on the initial velocity, which is neglected in the Gaussian distribution.
This dependence is slightly anisotropic being more pronounced in the direction perpendicular to the initial velocity than parallel to it.}
\end{figure}

We finally compare the distribution of the tagged particle, $n^\mathrm{tag}(\V{x},t)$, obtained from the MD simulations with the Gaussian distribution, $n^{\text{Gauss}}(\V{x},t)$, which is the solution of the diffusion equation~\eqref{PxtPx0}.
For a better testing, we correct for the small bias of $n^{\text{tag}}(\V{x},t)$ that emerges from the initial condition of the tagged particle having only velocities in the positive $x$-direction, by shifting the Gaussian density into the $x$-direction by the mean value $\langle x(t) \rangle$ of the tagged particle at the considered time $t$.
The latter is well described by \cite{PorraWangMasoliver1996,DespositoVinales2009} 
\begin{equation}
\label{avgxt}
\langle x(t)\rangle \approx \frac{d D(t)}{C(0)}\overline{v_0}.
\end{equation} 
This expression is obtained by integrating both sides of \eqref{VACFrel2} up to time $t$ and taking  the average over the absolute values of the thermally distributed initial velocities.   
As can be seen from Fig.~\ref{fig_ntr}, the marginal densities $n^\mathrm{tag}(x,t)\equiv\int n^\mathrm{tag}(x,y,t) dy$ according to the MD simulations and the shifted Gaussian density resulting from the diffusion equation~\eqref{PxtPx0} agree almost perfectly with each other.  
In conclusion, the tagged particle density is faithfully described by a Gaussian distribution with variance $2 \int D(t') dt'$ and mean values $\langle y(t) \rangle =0$ and $\langle x(t) \rangle$ given by \eqref{avgxt}.

\subsection{\label{subsec_vacf}Velocity Autocorrelation Function}

Based on the previous results, we also examine the validity of the analytic expression~\eqref{Ctlongtime} for the VACF $C(t)$. Its derivation is mainly based on the following three assumptions:  
\begin{itemize}
\item Assumption 1.\ According to \eqref{Ct1} and \eqref{Ct2}, $C(t)$ can be expressed in terms of the velocity field $\V{u}(\V{x},t)$, which is generated by the motion of the tagged particle relative to the fluid at the initial time, and the probability density $n^\mathrm{tag}(\V{x},t)$ to find the tagged particle at a later time $t$ at the position $\V{x}$.
\item Assumption 2.\ The velocity field $\V{u}(\V{x},t)$ can be obtained as a solution of the LNS.
\item Assumption 3.\ The velocity field $\V{u}(\V{x},t)$ can be substituted by its transverse component $\V{u}^\perp(\V{x},t)$.
\end{itemize}
We note here that Assumption~1 indicates that the tagged particle behaves as any other fluid particle. 
It still leaves open whether the velocity field $\V{u}(\V{x},t)$ is estimated from the positions and velocities of all fluid particles or whether it is obtained from a hydrodynamic consideration as postulated in Assumption~2.
Assumption~3 introduces a further simplification which though is expected to lead to errors at short times, and for finite systems also at large times. 

\begin{figure}
\centerfloat
\includegraphics[width=1.1\linewidth]{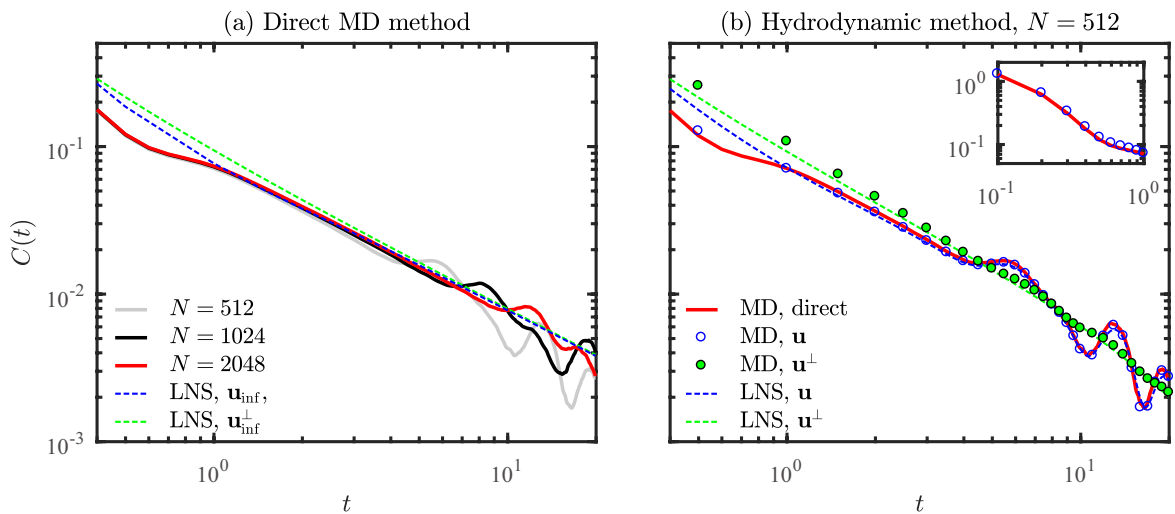}
\caption{\label{fig_vacf}
Comparison of VACFs $C(t)$ evaluated by various methods.
In panel~(a), the VACFs obtained directly from the MD simulations of three system sizes are depicted by solid lines.
The VACFs computed using the LNS solutions $\V{u}_\mathrm{inf}(\V{x},t)$ and $\V{u}_\mathrm{inf}^\perp(\V{x},t)$ for the infinite system limit are also plotted by dashed lines.
Hence, the green dashed line corresponds to \eqref{Ctlongtime}. 
In panel~(b), VACFs obtained from different methods for $N=512$ are compared.
The VACF obtained directly from MD is again displayed by the red solid line.
The VACFs computed from \eqref{Ct2} using the MD results $\V{u}$ and $\V{u}^\perp$ are depicted by blue (empty) and green (filled) circles, respectively.
Corresponding results obtained from \eqref{ksum} with the LNS solutions $\V{u}(\V{k},t)$ and $\V{u}^\perp(\V{k},t)$ are drawn by the blue and green dashed lines, respectively.
An extension to shorter times is depicted in the inset displaying a very good agreement of the directly calculated MD VACF and the result of \eqref{Ct2} with the full velocity field obtained from MD.   
The analogous results presented in panel (b) for $N=$1024 and 2048 do not provide further insights and therefore are not shown.}
\end{figure}

In Fig.~\ref{fig_vacf}~(a), we compare the VACFs obtained by the MD simulations of systems with three different sizes.
The expected $t^{-1}$ long-time tails emerge first around $t=1$ and well describe the VACFs up to the appearance of a series of humps with maxima at times $\sqrt{n^2_x+n^2_y} L/c_\mathrm{s}$ with $n_x,n_y = 0,1,2,\ldots$.
At these times a signal propagating with the sound-velocity $c_\mathrm{s}$ reappears when moving on a square with side length $L$ and periodic boundary conditions. Hence, the humps can be interpreted as the ``echoes'' of the initial state.
Actually, relatively insignificant deviations from the long-time tail behavior appear already before the first hump as was found from MD simulations, which are not presented here, of a much larger system with $N=6.5 \times 10^5$ particles.

In the intermediate time interval $5<t<L /c_\mathrm{s}$, the analytic expression~\eqref{Ctlongtime} being indicated by the green line in Fig.~\ref{fig_vacf}~(a) very well agrees with the MD result for the large system with $N=2048$.
Yet small but noticeable deviations are present in the time span $1<t<5$.
A much better conformity is achieved with the result calculated with the full solution $\V{u}(\V{x},t)$ of the LNS, see Appendix~\ref{appendix_LNS}, and the Gaussian distribution of the tagged particle (blue line in Fig.~\ref{fig_vacf}).
Hence, the deviations in the region $1<t<5$ can be fully attributed to a violation of Assumption~3. 

In order to assess the finite-size effects in particular in combination with further approximations, we also considered the finite size version of \eqref{Ct2}, which is of the form
\begin{equation}
\label{ksum}
C(t) \approx  \frac{C(0)}{\langle v_0\rangle L^d}\sum_{\V{k}\ne\V{0}} u_x(-\V{k},t)n^\mathrm{Gauss}(\V{k},t).
\end{equation}
Here, a cut-off $k_\mathrm{c} = \pi/\D{x}_\mathrm{c}$ at large $k$ values must be introduced in order to avoid the divergence   of $C(t)$ at $t=0$.~\cite{Kawasaki1970b,KeyesLadanyi1975}
We choose as cut-off length $\D{x}_\mathrm{c}=0.1$, in agreement with the spatial resolution used for the estimation \eqref{fieldest} of the MD fields.  

In Fig.~\ref{fig_vacf}~(b), the VACF directly obtained from the MD simulations is compared with the result from \eqref{Ct2} with the full velocity field obtained from MD simulations (blue dots) as well as the result from \eqref{ksum} with the LNS $\V{u}(\V{x},t)$ of the corresponding finite system with periodic boundary conditions (blue dashed line). 
The LNS equations are solved with time-dependent transport coefficients based on molecular expressions, see Appendix~\ref{appendix_GreenKubo}. 
Whether these transport coefficients and the thermodynamic parameters listed in Appendix~\ref{appendix_thermo} are determined in finite-size systems or in the limit of an infinite-size system turns out to be insignificant.

While the result from the full MD velocity field perfectly agrees with the direct MD VACF from the shortest to the largest times, the LNS result coincides at short and intermediate times with the infinite system results hence revealing the aforementioned deviations at short times.
At large times, the humps caused by echoes on the finite torus are perfectly reproduced by \eqref{ksum} provided the full velocity $\V{u}(\V{x},t)$ is used, whereas with a transversal velocity field $\V{u}^\perp(\V{x},t)$ these humps in $C(t)$ disappear independent of whether the respective MD (green dots) or the LNS velocity field (green dashed line) is used.

In view of the above formulated assumptions, we may conclude that Assumption~1 is surprisingly well satisfied even in the kinematic region with $t <1$, where the LNS predictions fail mainly for the longitudinal components as evident from the previous Figs.~\ref{t05}~(c), (f), \ref{fig_uk}~(c), (d), and \ref{fig_upar}.
Therefore, Assumption~2 cannot be used for times $t<1$, which corresponds to less than 5 mean collision times between fluid particles. 
For $t \gtrsim 1$ though, the full LNS velocity field yields a valid description of $C(t)$ for all times, including the echo humps of finite systems; hence, Assumption~2 is justified from the time $t =1$ on.  
Assumption~3 sets in to hold for times $t \gtrsim 5$.
This time scale is with approximately 25 mean collision times still extremely short if considered at the hydrodynamic time scale.
Finite size effects are mainly suppressed with this assumption, which therefore well describes the thermodynamic limit.

\section{\label{sec_summary}Summary and Conclusion}

In the present study, we investigated self-diffusion in a system of $N$ pairwise interacting soft disks moving in a two-dimensional square with periodic boundary conditions and compared with the results predicted by the linearized Navier--Stokes equations.
By computing hydrodynamic field variables  directly from extensive MD simulations with high spatial and temporal resolution and by comparing with the solutions of the LNS, we performed a quantitative analysis indicating the regime of validity of the LNS approach both with respect to the length and time scales.
In particular, in view of the fact that certain transport coefficients such as the diffusion coefficient do not exist for two-dimensional fluids, it was of significant importance to use time-dependent transport coefficients in the LNS.
These time-dependent transport coefficients are determined by finite-time Green--Kubo relations.
We expect that their use may also improve the agreement of the solutions of the LNS and results from MD simulations for three-dimensional systems at small spatial and short time scales, and in this way may narrow the gap between the kinetic and the field theoretic descriptions of a fluid.

In the presently investigated two-dimensional case, we were able to completely close this gap and demonstrate that the hydrodynamic description is valid already after a few mean collision times.
The predictions of the LNS set in to hold already at very short times for the perpendicular part of the velocity field, $\V{u}^\perp(\V{x},t)$, that contains the vortex pair eventually being responsible for the long-time tail of the VACF.
On the other hand, the longitudinal wave-like part of the velocity field, $\V{u}^\parallel(\V{x},t)$, resulting from the MD simulations contains at short times features at molecular scales that are absent in the respective LNS field.
The layered spatial structure of the MD longitudinal velocity field $\V{u}^\parallel(\V{x},t)$ at short times is caused by the typical shell-like structure of the radial distribution function.     
The asymmetry of the forward and backward wavefronts can be explained by the backscattering events of the tagged particle necessary to build up the backward oriented wave front.
However, by comparing the growth of flow patterns, we confirmed that outside the kinetic region the LNS solutions provide accurate quantitative description for both $\V{u}^\perp(\V{x},t)$ and $\V{u}^\parallel(\V{x},t)$.

Based on these observations, we revisited the long-time decay of the VACF, which was first observed and explained by Alder and Wainwright and subsequently analyzed by the mode-coupling theory.
We investigated the implicit assumptions of the mode-coupling theory and specified the underlying assumptions as well as their applicability.
As the central results, we found that the description of the VACF in terms of (i) a fluid velocity field $\V{u}(\V{x},t)$ that is conditioned on the initial position and velocity of the tagged particle, and (ii) the probability density $n^{\text{tag}}(\V{x},t)$ of the tagged particle, quantitatively agrees with the standard definition as the velocity-velocity correlation function of a selected fluid particle from very short times on, which are less than the mean collision time.   
The replacement of the named velocity field by the solution of the LNS equations with properly determined time-dependent transport coefficients is restricted to times that must be of the order of five mean collision times or larger and is hence already valid on quite short time scales compared to standard hydrodynamic times. 
The use of time-dependent transport coefficients is mandatory because of the divergence of transport coefficients in two-dimensional fluids, but may also improve the solutions of three-dimensional fluids at short times. 

The deviations of the actual flow patterns from those resulting from LNS at very short times can be traced back to the atomistic structure of the fluid but also might be influenced by the nonlinearity of the full Navier--Stokes equations, which at short times must not be neglected in view of the singular initial condition~\eqref{ux0}.
Finally,  the velocity field may be replaced by its transversal part after the relatively short time that it takes until the wave-like propagating part of the velocity does no longer overlap with the tagged particle probability density. 
The transversal velocity part contains in two dimensions a vortex pair and in three dimensions a vortex ring.
These structures  are responsible for the algebraically slow decay of the VACF as it already was predicted by Alder and Wainwright. 

The expression~\eqref{Ctlongtime} relating the VACF and the diffusion coefficient to each other in a nonlinear way was recently used by some of the present authors to find the large time behavior of these quantities resulting in proper corrections to the $t^{-1}$ decay law of $C(t)$ and the conforming diffusion coefficient $D(t)$ for a two-dimensional fluid.~\cite{ChoiHanKimTalknerKideraLee}  

A phenomenon closely related to self-diffusion to which the present approach might be applied is Brownian motion.~\cite{BianKimKarniadakis2016, ShinKimTalknerLee2010}
In the case of a Brownian particle, the considered  particle has different, mostly much larger mass and also a larger volume than a fluid particle. While in the Brownian limit of heavy particles normal diffusion governed by a Markovian Langevin equation characterizes the dynamics well, the cross-over behavior in a two-dimensional fluid from anomalous self-diffusion to normal diffusion presents an open problem.

\begin{acknowledgments}
This work was supported by the U.S. Department of Energy, Office of Science, Office of Advanced Scientific Computing Research, Applied Mathematics Program under Contract No.~DE-AC02-05CH11231, and Korea Advanced Institute of Science and Technology (KAIST), College of Natural Science, Research Enhancement Support Program under Grant No.~A0702001005.
An award of computer time was provided by the ASCR Leadership Computing Challenge (ALCC) program.
This research used resources of the Oak Ridge Leadership Computing Facility, which is a DOE Office of Science User Facility supported under Contract No.~DE-AC05-00OR22725.
\end{acknowledgments}

\appendix

\section{\label{appendix_LNS}Linearized Navier--Stokes Equations}

We present the form of the LNS equations considered in the paper and obtain analytic results \eqref{vortexy0} for the transverse velocity in the two-dimensional case.
For the number density $n(\V{x},t)$, the velocity $\V{u}(\V{x},t)$, and the temperature $T(\V{x},t)$, the linearized governing equations of a compressible fluid are written as~\cite{ErnstHaugeLeeuwen1971b,BoonYip1980,HansenMcDonald2013}
\begin{subequations}
\label{LNS}
\begin{align}
\label{LNS1}
& \frac{\partial n(\V{x},t)}{\partial t} = -\bar{n} \divg\V{u}(\V{x},t), \\
\label{LNS2}
& \frac{\partial \V{u}(\V{x},t)}{\partial t}= -\frac{c_\mathrm{s}^2}{\gamma\bar{n}}\grad n(\V{x},t) + \nu \lapl \V{u}(\V{x},t) + (D_{\mathrm{l}}-\nu)\grad(\divg\V{u}(\V{x},t))-\frac{c_\mathrm{s}^2\alpha}{\gamma}\grad T(\V{x},t), \\
\label{LNS3}
& \frac{\partial T(\V{x},t)}{\partial t}= -\frac{\gamma-1}{\alpha}\divg\V{u}(\V{x},t)+\gamma D_{\mathrm{T}} \lapl T(\V{x},t),
\end{align}
\end{subequations}
where $\bar{n}$ is the mean number density, $c_\mathrm{s}$ is the adiabatic speed of sound, $\gamma$ is the ratio of specific heats, $\nu$ is the kinematic viscosity, $D_\mathrm{l}$ is the kinematic longitudinal viscosity, $\alpha$ is the thermal expansion coefficient, and $D_\mathrm{T}$ is the thermal diffusivity.

The solution of \eqref{LNS} obeying the initial conditions \eqref{ux0} and \eqref{nx0Tx0} is readily expressed in a closed form in the Fourier space.
By assuming time-dependent coefficients $\nu(t)$, $D_\mathrm{l}(t)$, and $D_\mathrm{T}(t)$ and $\V{v}_0 = v_0 \V{e}_x$, the longitudinal and transverse velocities $\tilde{\V{u}}^\bullet(\V{k},t) = \int d \V{x}\: e^{-i \V{k} \cdot \V{x}} \V{u}^\bullet(\V{x},t)$, $\bullet = \perp,\; \parallel$, become
\begin{subequations}
\label{ukt}
\begin{align}
\label{uperpkt}
&\tilde{\V{u}}^{\perp}(\V{k},t)=\frac{v_0}{\bar{n}}\left[\V{e}_x-\frac{k_x}{k}\hat{\V{k}}\right]e^{-k^2\int_0^t\nu(t')dt'},\\
\label{uparallelkt}
&\tilde{\V{u}}^{\parallel}(\V{k},t)=\frac{v_0}{\bar{n}}\frac{k_x}{k}\hat{\V{k}}\cos{(c_\mathrm{s} k t)} e^{-\frac{k^2}{2}\int_0^t\Gamma_{\mathrm{s}}(t')dt'},
\end{align}
\end{subequations}
where $k$ and $\hat{\V{k}}$ denote the magnitude and unit vector of $\V{k}$, respectively, and $\Gamma_{\mathrm{s}}(t)=D_{\mathrm{l}}(t)+(\gamma-1)D_{\mathrm{T}}(t)$ is the sound attenuation coefficient.

In the two-dimensional case, from the inverse Fourier transform of \eqref{uperpkt}, the transverse velocity has the following closed-form expression 
\begin{subequations}
\label{uperprthetat}
\begin{align}
&u^{\perp}_x(r,\theta,t) = \frac{v_0}{4\pi\bar{n}r^2} \left( 2\cos 2\theta - \frac{2B(t) \cos 2\theta - r^2\sin^2\theta}{B(t)} e^{-\frac{r^2}{4B(t)}} \right), \\
&u^{\perp}_y(r,\theta,t) = \frac{v_0}{8\pi\bar{n}r^2} \left( 4 - \frac{4B(t)+r^2}{B(t)} e^{-\frac{r^2}{4B(t)}} \right)\sin 2\theta,
\end{align}
\end{subequations}
where $r$ and $\theta$ denote the polar coordinates of $\V{x}$ and $B(t)=\int_0^t\nu(t')dt'$.
From the condition $\V{u}^\perp=\V{0}$, we can easily find the location of each vortex center $(0,\pm y_0)$, where 
\begin{equation}
\label{vortexy0}
y_0 = 2\sqrt{\xi B(t)} \approx 2.24 \Big[\textstyle\int_0^t \nu(t')dt'\Big]^{1/2}
\end{equation}
and $\xi\approx 1.26$ is the positive solution of $(\xi+\frac12)e^{-\xi}=\frac12$.
For convergent $\nu(t)$ (i.e., $\lim_{t\rightarrow\infty}\nu(t)=\nu$), we have $\int_0^t\nu(t')dt'\approx\nu t$ at large $t$ and thus obtain the $\sqrt{\nu t}$ time scale of $y_0$.

\section{\label{appendix_GreenKubo}Transport Coefficients from Green--Kubo formulas}

\begin{figure}
\centerfloat
\includegraphics[width=1.1\linewidth]{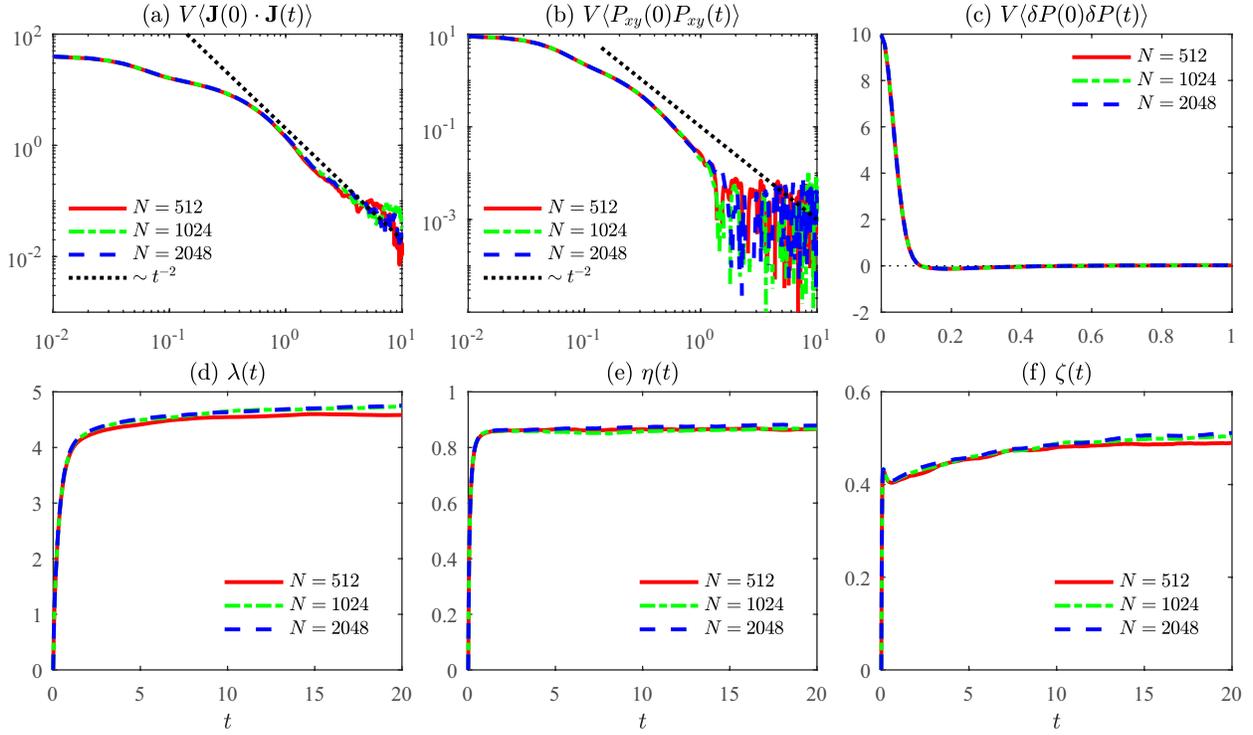}
\caption{\label{acf}
MD estimation of transport coefficients. The time-dependent behavior of the autocorrelation functions for $\V{J}(t)$, $P_{xy}(t)$, and $\delta P(t)$ are presented in panels~(a), (b), and (c), respectively, and the corresponding transport coefficients are shown in panels~(d), (e), and (f).
The red solid lines, green dashed-dot lines, and blue dashed lines depict the results of $N=512$, 1024, and 2048, respectively ($V=L^2=853.3$, 1707, and 3413).
The black dotted lines represent auxiliary lines corresponding to $t^{-2}$.}
\end{figure}

We consider the following five time-dependent transport coefficients which are defined by Green--Kubo relationship in terms of correlation functions:
\begin{subequations}
\begin{align}
&\mbox{Diffusion coefficient: } D(t) = \frac{1}{d} \int^t_0 \langle\V{v}(0)\cdot\V{v}(t) \rangle dt', \\
&\mbox{Heat conductivity: }     \lambda(t) = \frac{V}{dk_\mathrm{B}T^2} \int^t_0 \langle \V{J}(0)\cdot\V{J}(t')\rangle dt', \\
&\mbox{Shear viscosity: }       \eta(t) = \frac{V}{k_\mathrm{B}T} \int^t_0 \langle P_{xy}(0)P_{xy}(t')\rangle dt',\\
&\mbox{Bulk viscosity: }        \zeta(t) = \frac{V}{k_\mathrm{B}T} \int^t_0 \langle \delta P(0)\delta P(t')\rangle dt',\\
&\mbox{Kinematic viscosity: }   \nu(t) = \frac{\eta(t)}{\bar{n}m}.
\end{align}
\end{subequations}
Then the respective flux are defined as
\begin{subequations}
\begin{align}
\V{J}&=\displaystyle\frac{1}{V}\left[\displaystyle\sum_i{e_i{\V{v}}_i}+\frac{1}{2}\displaystyle\sum_{j\neq i}{\left(\V{f}_{ij}\cdot \V{v}_j\right)\V{r}_{ij}}\right], \\
P_{xy}&=\displaystyle\frac{1}{V}\left[\displaystyle\sum_i mv_{ix}v_{iy}+\frac{1}{2}\displaystyle\sum_{j\neq i}{r_{ijx}f_{ijy}}\right],\\
\delta P&=\displaystyle\frac{1}{dV}\left[\displaystyle\sum_i m\V{v}_i\cdot\V{v}_i +\frac{1}{2}\displaystyle\sum_{j\neq i}{\V{r}_{ij}\cdot \V{f}_{ij}}\right] -\left\langle P\right\rangle.
\end{align}
\end{subequations}
Here $V$ and $\left\langle P\right\rangle$ are the volume and the average pressure of the system, respectively, and $e_i$, $\V{f}_{ij}$ and $\V{r}_{ij}$ are the energy of atom $i$, the inter-atomic force, and position vector between atoms $i$ and $j$, respectively.
Figure~\ref{acf} displays the MD estimation results of these flux correlation functions and the respective transport coefficients.

The sound attenuation coefficient, defined as
\begin{equation}
\Gamma_\mathrm{s}(t)=D_\mathrm{l}(t)+(\gamma-1)D_\mathrm{T}(t),
\end{equation}
can be obtained from the heat conductivity $\lambda(t)$, shear viscosity $\eta(t)$, and bulk viscosity $\zeta(t)$, whereby the thermal diffusivity $D_\mathrm{T}(t)$ and the longitudinal diffusivity (kinematic longitudinal viscosity) $D_\mathrm{l}(t)$ are related to these three coefficients as
\begin{subequations}
\begin{align}
D_\mathrm{T}(t)& =\frac{\lambda(t)}{\bar{n}C_P}, \\
D_\mathrm{l}(t)& = \frac{1}{\bar{n}m}\left(\frac{2\left(d-1\right)}{d}\eta (t)+\zeta (t)\right),
\end{align}
\end{subequations}
where $C_P$ and $\gamma$ are the isobaric heat capacity and the ratio of the specific heats of the system, respectively.

\section{\label{appendix_thermo}Thermodynamic Parameters from the Phase Space Volume}

Here we briefly review Meier and Kabelac’s method~\cite{MeierKabelac2006} and apply this method for the calculation of the necessary thermodynamic parameters in  two-dimensional fluid systems.  
Starting from the entropy $S=k_\mathrm{B}\ln\Omega$, where $\Omega$ is the phase-space volume, the adiabatic speed of sound, for example, defined as  
\begin{equation}
c_\mathrm{s}=\sqrt{{\left(\frac{\partial P}{\partial \rho }\right)}_S} =\sqrt{-\frac{V^2}{M}{\left(\frac{\partial P}{\partial V}\right)}_S},
\end{equation}
can be expressed in terms of derivatives of the phase space volume with respect to the energy $E$ and the volume $V$, yielding
\begin{equation}
c_\mathrm{s}=\sqrt{\frac{V^2}{M}\left(\frac{1}{\omega}\frac{\partial\mathrm{\Omega}}{\partial V}\left(2\frac{1}{\omega}\frac{\partial^{2}\Omega}{\partial V\partial E}-\frac{1}{{\omega }^{2}}\frac{{\partial }^{2}\mathrm{\Omega}}{\partial E^{2}}\frac{\partial \Omega }{\partial V}\right)-\frac{1}{\omega }\frac{\partial^{2}\Omega}{\partial V^{2}}\right)},
\end{equation} 
where $\omega=\frac{\partial \Omega}{\partial E}$ is the phase-space density, $M$ and $\rho$ are the total mass (i.e., $M=\sum_{i=1}^{N}m_i$) and mass density (i.e., $\rho=M/V$) of the system, respectively.

Defining the center of mass related quantity $\V{G}$
\begin{equation}
\V{G}=\sum^N_{i=1}{\V{p}_i t}-\sum^N_{i=1}{m_i\V{x}_i}=\V{P}t-\sum^N_{i=1}{m_i\V{x}_i}, 
\end{equation}
we express the phase space volume depending on the conserved quantities $N$, $V$, $E$, $\V{P}$, $\V{G}$, as 
\begin{equation}
\label{Omega}
\begin{split}
\Omega\left(N,V,E,\V{P},\V{G}\right)&=\frac{1}{C_N}\int d \V{x}^N \int d{\V{p}}^N \Theta \left(E-\frac{\V{P}\cdot\V{P}}{2M}-H\right) \\
&\times\delta\left(\V{P}-\sum^N_{i=1}{\V{p}_i}\right)\delta \left(\V{G}-t\sum^N_{i=1}{\V{p}_i}+\sum^N_{i=1}{m_i\V{x}_i}\right), 
\end{split}
\end{equation} 
where $\V{P}=\sum^N_{i=1}{\V{p}_i}$ is the total momentum, $H=K+U$ the Hamiltonian of the system, $K$ the kinetic energy of the system, $U$ the potential energy of the system, $\Theta$ the Heaviside step function, and $C_N$ a normalization constant rendering the phase space volume dimensionless. 

\begin{table}
\centering
\begin{tabular}{l}
\hline \hline
$\Omega_{00}=\frac{1}{N-1}\left\langle K\right\rangle \mathrm=k_{\mathrm{B}}T$ \\ 
$\Omega_{10}=1$ \\ 
$\Omega_{01}=\frac{1}{V}\left\langle K\right\rangle -\left\langle \frac{\partial U}{\partial V}\right\rangle =P$ \\ 
$\Omega_{11}=\frac{N-1}{V}-\left(N-2\right)\left\langle K^{-1}\left(\frac{\partial U}{\partial V}\right)\right\rangle$ \\ 
$\Omega_{20}=\left(N-2\right)\left\langle K^{-1}\right\rangle$ \\
$\Omega_{02}=\frac{N-2}{V^2}\left\langle K\right\rangle -2\frac{N-1}{V}\left\langle \frac{\partial U}{\partial V}\right\rangle -\left\langle \frac{{\partial }^2U}{\partial V^2}\right\rangle +\left(N-2\right)\left\langle K^{-1}{\left(\frac{\partial U}{\partial V}\right)}^2\right\rangle$  \\
\hline \hline
\end{tabular}
\caption{\label{phasespace}
The phase-space functions ${\Omega}_{mn}$ $(m+n\leq2)$ of the two-dimensional $NVE\V{P}\V{G}$ ensemble.}
\end{table}

We define the phase-space functions as $\Omega_{mn}=\frac{1}{\omega }\frac{{\partial }^{m+n}\Omega}{\partial E^m\partial V^n}$.
Expressions of ${\Omega}_{mn}$ in terms of the kinetic energy $K$ and the volume derivatives of the potential energy $\partial^n U/\partial V^n$ can be derived by using derivatives of \eqref{Omega}.
We list the phase-space functions ${\Omega}_{mn}$ $(m+n\leq2)$ of the two-dimensional $NVE\V{P}\V{G}$ ensemble in Table~\ref{phasespace}.

Thermodynamic state functions appearing in \eqref{LNS1} and \eqref{LNS2},
\begin{subequations}
\begin{align}
&\mbox{Adiabatic speed of sound: } c_\mathrm{s}=\sqrt{\frac{V}{M\beta_{S}}}, \\
&\mbox{Thermal expansion coefficient: } \alpha ={\beta }_T{\gamma }_V, \\ 
&\mbox{Ratio of the specific heats: } \gamma =\frac{C_P}{C_V}, \\
&\mbox{Isobaric heat capacity: } C_P=C_V\frac{{\beta }_T}{{\beta }_S},
\end{align}
\end{subequations}
can be expressed in terms of ${\mathrm{\Omega }}_{mn}$ using
\begin{subequations}
\begin{align}
&\mbox{Isochoric heat capacity: } C_V=k_\mathrm{B}{\left(1-{\Omega}_{00}{\Omega}_{20}\right)}^{-1},\\
&\mbox{Thermal pressure coefficient: } \gamma_V=k_\mathrm{B}\frac{\Omega_{11}-\Omega_{01}\Omega_{20}}{1-\Omega_{00}\Omega_{20}}, \\
&\mbox{Isothermal compressibility: } \beta^{-1}_T=V\left[\frac{\Omega_{01}\left(2\Omega_{11}-\Omega_{01}\Omega_{20}\right)-\Omega_{00}\Omega^2_{01}}{1-\Omega_{00}\Omega_{20}}-\Omega_{02}\right],\\
&\mbox{Adiabatic compressibility: } \beta^{-1}_S=V\left[\Omega_{01}\left(2\Omega_{11}-\Omega_{01}\Omega_{20}\right)-\Omega_{02}\right].
\end{align}
\end{subequations}

\section{\label{appendix_Gal}Objective Vortex Identification}

\begin{figure}
\centerfloat
\includegraphics[width=0.55\linewidth]{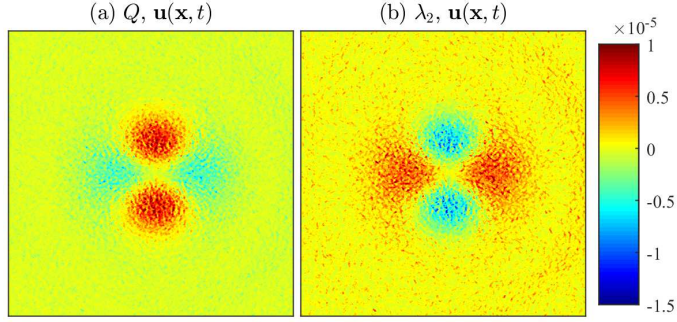}
\caption{\label{fig_vorticity}
Test of two objective criteria for vortices.
According to the $Q$- and the $\lambda_2$-criteria, two vortices can be consistently identified located in regions around the zeroes of the transversal fields $\V{u}^\perp(\V{x},t)$. This velocity field is obtained for a system of size $N=2048$ at time $t=4$.
The displayed spatial domain is restricted to  $\V{x}\in(-15,15)\times(-15,15)$.}
\end{figure}

In order to demonstrate that the pair of vortices that are obtained by means of the Helmholtz decomposition of the MD vector field is Galilean-invariant, we use the following two objective criteria.~\cite{HuntWrayMoin1988,JeongHussain1995,Haller2005}
Both criteria rely on the gradient $\V{\nabla} \V{u}$ of the considered velocity field $\V{u}$, which is decomposed into its symmetric and anti-symmetric parts $\V{S}$ and $\V{A}$, respectively, defined as
\begin{equation}
\begin{split}
\V{S} &= \frac{1}{2} \left [ \V{\nabla}\V{u} +(\V{\nabla}\V{u})^T \right], \\
\V{A} &= \frac{1}{2} \left [ \V{\nabla}\V{u} -(\V{\nabla}\V{u})^T \right],
\end{split}
\end{equation}
where $\V{X}^T$ denotes the transpose of the matrix $\V{X}$.
 
The first, so-called $Q$-criterion locates a vortex where the scalar function $Q$ defined as
\begin{equation}
Q=\frac{1}{2}\left[|\V{\Omega}|^2-|\V{S}|^2\right]
\end{equation}
is positive.
Here $|\V{X}|^2 = \mathrm{Tr} (\V{X} \V{X}^T)$ defines a norm of the matrix $\V{X}$.
The second, so-called $\lambda_2$-criterion locates a vortex where the second eigenvalue $\lambda_2$ of the matrix $\V{S}^2 + \V{\Omega}^2$ is negative, whereby the two eigenvalues of  $\V{S}^2 + \V{\Omega}^2$ are ordered such that $\lambda_1 > \lambda_2$.   

Figure~\ref{fig_vorticity} confirms that there are two vortices with positions located around $(0,\pm y_0)$ with $y_0$ defined in \eqref{vortexy0}.

\bibliographystyle{aipnum4-1}
\bibliography{main}

\end{document}